 \documentclass[manuscript]{emulateapj-rtx4}

\newcommand{\rhoGJ}{\rho_{{\rm GJ}}}  
\newcommand{\Ell}{E_\parallel}      
\newcommand{\rhowSQR}{\rho_{\rm w}^2}

\slugcomment{}

\shorttitle{Gamma Radiation from Rotating Black Holes}
\shortauthors{Hirotani and Pu}

\begin{document}


\title{Energetic Gamma Radiation from Rapidly Rotating Black Holes}


\author{Kouichi Hirotani, and Hung-Yi Pu}
\affil{Academia Sinica, Institute of Astronomy and Astrophysics (ASIAA),
       PO Box 23-141, Taipei, Taiwan}
\email{hirotani@tiara.sinica.edu.tw}

%
%
%
%


\begin{abstract}
Supermassive black holes are believed to be the central power house 
of active galactic nuclei. 
Applying the pulsar outer-magnetospheric particle accelerator theory
to black-hole magnetospheres, 
we demonstrate that an electric field is exerted along the magnetic 
field lines near the event horizon of a rotating black hole.
In this particle accelerator (or a gap), 
electrons and positrons are created by photon-photon collisions
and accelerated in the opposite directions by this electric field,
efficiently emitting
gamma-rays via curvature and inverse-Compton processes.
It is shown that a gap arises around the null charge surface
formed by the frame-dragging effect,
provided that there is no current injection across the gap boundaries.
The gap is dissipating a part of the hole's rotational energy, and 
the resultant gamma-ray luminosity increases 
with decreasing plasma accretion from the surroundings. 
Considering an extremely rotating supermassive black hole, 
we show that such a gap reproduces the significant very-high-energy 
(VHE) gamma-ray flux observed from the radio galaxy IC 310,
provided that the accretion rate becomes much less
than the Eddington rate particularly during its flare phase.
It is found that the curvature process dominates the
inverse-Compton process in the magnetosphere of IC~310,
and that the observed power-law-like spectrum in VHE 
gamma-rays can be explained to some extent by a superposition of 
the curvature emissions with varying curvature radius.
It is predicted that the VHE spectrum extends into higher energies
with increasing VHE photon flux.
\end{abstract}





\section{Introduction}
%
%
The MAGIC (Major Atmospheric Gamma-ray Imaging Cherenkov) telescopes
recently reported on gamma-ray observations of the radio galaxy IC 310
\citep{alek14b}, 
which presumably harbors a black hole (BH) 
with 0.1-0.7 billion solar masses 
\citep{alek14b,macElroy95,simien02}. 
MAGIC detected a powerful flare in very high energies (VHE) between 
70 GeV and 8 TeV and revealed that the flare flickers with a doubling 
time scale as short as 4.8 minutes. 
Ruling out other possibilities, the MAGIC team inferred that a substantial
portion of the flaring photons originated from a compact region 
as small as one fifth of the typical BH radius, or the 
Schwarzschild radius, $2 r_{\rm g}=3 \times 10^{14} M_9$ cm, 
where $M_9= M/(10^9 M_\odot)$ denotes 
the BH mass in billion solar masses.

To interpret this amazing finding of a unique sub-horizon phenomenon, 
we apply the pulsar vacuum-gap model 
\citep{cheng86a} to the BH magnetosphere 
of IC 310,
extending the method developed by 
\citet{bes92} and \citet[hereafter HO98]{hiro98}. 
It is known that the rotational energy of a BH can be 
electromagnetically extracted with spin-down luminosity 
\citep{bla76,koide02}
\begin{equation}
  L_{\rm sd} 
  = \Omega_{\rm F} (\omega_{\rm H}-\Omega_{\rm F})
    B_\perp{}^2 r_{\rm H}{}^4 / c, 
  \label{eq:Lsd}
\end{equation}
where $\Omega_{\rm F}$ denotes the angular frequency of the rotating 
magnetic field lines, 
$\omega_{\rm H}= ac/[2(GMc^{-2})r_{\rm H}]$ 
the angular frequency of the rating BH,
$B_\perp$ the normal component of the magnetic field threading 
the event horizon, 
$r_{\rm H}$ the radius of the horizon, and 
$c$ the speed of light. 
A portion of such a spin-down energy can be dissipated 
as photon emissions within the magnetosphere.

In \S~\ref{sec:nullS}, 
we outline previous BH gap models,
arranging their electrodynamic properties (and hidden philosophies)
according to the Poisson equation for the non-corotational potential.
Then in \S~\ref{sec:RIAF},
we point out that the magnetosphere of IC~310 is usually
filled with copious plasmas that quench a BH gap,
and that a BH gap can be switched on if the mass accretion rate
is halved from the time-averaged value.
We formulate the BH gap in \S~\ref{sec:phys},
and apply the model to IC~310 in \S~\ref{sec:app_IC310}.
We highlight the difference among current BH gap models
in the final section.

\section{Gap position}
\label{sec:gap_position}
In this section, we investigate the plausible position
of particle accelerators in a black hole magnetosphere.

\subsection{The null charge surface}
\label{sec:nullS}
When magnetized plasmas are accreting onto an astrophysical BH, 
the self-gravity of the plasma particles and 
the electromagnetic field little affects the space-time geometry.
Thus, around a rotating BH, 
the background geometry is described by the Kerr metric
\citep{kerr63}.
In the Boyer-Lindquist coordinates, it becomes 
\citep{boyer67} 
\begin{equation}
 ds^2= g_{tt} dt^2
      +2g_{t\varphi} dt d\varphi
      +g_{\varphi\varphi} d\varphi^2
      +g_{rr} dr^2
      +g_{\theta\theta} d\theta^2,
  \label{eq:metric}
\end{equation}
where 
\begin{equation}
   g_{tt} 
   \equiv 
   -\frac{\Delta-a^2\sin^2\theta}{\Sigma} c^2 ,
   \qquad
   g_{t\varphi}
   \equiv 
   -\frac{2(GM/c^2) ar \sin^2\theta}{\Sigma} c, 
  \label{eq:metric_2}
\end{equation}
\begin{equation}
   g_{\varphi\varphi}
     \equiv 
     \frac{A \sin^2\theta}{\Sigma} , 
     \qquad
   g_{rr}
     \equiv 
     \frac{\Sigma}{\Delta} , 
     \qquad
   g_{\theta\theta}
     \equiv 
     \Sigma ;
  \label{eq:metric_3}
\end{equation}
$\Delta \equiv r^2-2(GM/c^2)r+a^2$,
$\Sigma\equiv r^2 +a^2\cos^2\theta$,
$A \equiv (r^2+a^2)^2-\Delta a^2\sin^2\theta$,
and $a \equiv J/(Mc)$,
$c$ denotes the speed of light, $M$ the mass of the BH,
$J$ the BH's angular momentum.
At the event horizon, $\Delta$ vanishes, giving
$r_{\rm H} \equiv r_{\rm g}+\sqrt{r_{\rm g}{}^2-a^2}$
as the horizon radius, where $r_{\rm g} \equiv GM c^{-2}$. 

In the same manner as in pulsar magnetospheres,
the Gauss's law, $\nabla_\mu F^{t\mu}=(4\pi/c)\rho$,
gives the Poisson equation 
for the non-corotational potential $\Psi$ \citep{hiro06},
\begin{equation}
  -\frac{c^2}{\sqrt{-g}}
   \partial_\mu 
      \left( \frac{\sqrt{-g}}{\rhowSQR}
             g^{\mu\nu} g_{\varphi\varphi}
             \partial_\nu \Psi
      \right)
  = 4\pi(\rho-\rhoGJ)
  \label{eq:pois}
\end{equation}
where the Greek indices run over $t$, $r$, $\theta$, $\varphi$,
$\sqrt{-g}= \sqrt{g_{rr}g_{\theta\theta}\rhowSQR}=c\Sigma\sin\theta$, 
$\rhowSQR \equiv g_{t\varphi}^2-g_{tt}g_{\varphi\varphi}$, and
$\rho$ the real charge density;
the general relativistic GJ charge density
is defined as \citep{GJ69,mestel71,hiro06}
\begin{equation}
  \rhoGJ \equiv 
      \frac{c^2}{4\pi\sqrt{-g}}
      \partial_\mu \left[ \frac{\sqrt{-g}}{\rhowSQR}
                         g^{\mu\nu} g_{\varphi\varphi}
                         (\Omega_{\rm F}-\omega) F_{\varphi\nu}
                 \right] .
  \label{eq:rhoGJ}
\end{equation}
The magnetic-field-aligned electric field can be computed by
$ \Ell \equiv (\mbox{\boldmath$B$}/B)
              \cdot \mbox{\boldmath$E$}
       = (\mbox{\boldmath$B$}/B)
              \cdot (-\nabla\Psi) .
  \label{eq:Ell}
$

If $\rho$ deviates from $\rhoGJ$ in any region,
$\Ell$ is exerted there along $\mbox{\boldmath$B$}$.
Once $E_\parallel$ arises, positive and negative charges migrate 
in opposite directions. 
For the magnetosphere to be force-free outside the gap, 
$\rho$ should match $\rho_{\rm GJ}$ at the boundaries. 
Since $\rho$ has opposite signs at the two boundaries, 
$E_\parallel$ should arise around the null charge surface,
where $\rho_{\rm GJ}$ vanishes, in the same way as pulsars
\citep{cheng00,romani96}. 
On these grounds, we consider that the null surface
is a natural place for a particle accelerator to arise.

\begin{figure}
 \epsscale{1.0}
 \includegraphics[angle=0,scale=0.43]{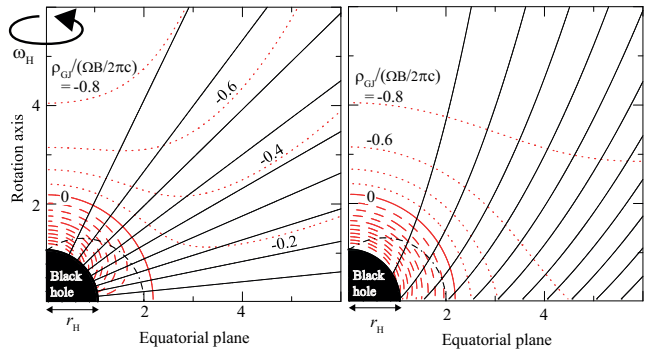}
\caption{
Distribution of the null surface (red thick solid curve)
on the poloidal plane in the Boyer-Lindquist coordinates.
The axes are in $r_{\rm g}=GMc^{-2}$ unit. 
The black hole (filled black region in the lower left corner)
rotates rapidly with spin parameter a=0.998 $r_{\rm g}$ around the ordinate. 
The contours of the dimensionless Goldreich-Julian charge density are 
plotted with the red dashed curves (for positive values) and 
the red dotted ones (for negative values as labeled). 
The black dash-dotted curve denotes the static limit, within which the 
rotational energy of the hole is stored. 
The black solid curves denote the magnetic field lines.
The left panel is the result for the case of a split-monopole
magnetic field, and the right one for a parabolic field. 
\label{fig:side_null}
}
\end{figure}

In figure~\ref{fig:side_null},
we plot the distribution of the null surface 
as the thick red solid curve.
In the left panel, we assume that the magnetic field is
split-monopole, adopting 
$A_\varphi \propto -\cos\theta$
as the magnetic flux function \citep{michel73}.
In the right panel, on the other hand,
we assume a parabolic magnetic field line
with $A_\varphi \propto r(1-\cos\theta)$
on the poloidal plane (i.e., $r$--$\theta$ plane).
We adopt $\Omega_{\rm F}=0.3 \omega_{\rm H}$ for both cases
\citep{mckinney12,bes13}. 
It follows that the null surface 
distributes nearly spherically, 
irrespective of the poloidal magnetic field configuration.
This is because its position is essentially determined by
the condition $\omega=\Omega_{\rm F}(A_\varphi)$,
because $\omega$ has weak dependence on $\theta$
near the horizon,
and because we assume $\Omega_{\rm F}$ is constant for
$A_\varphi$ for simplicity.
If $\Omega_{\rm F}$ decreases (or increases) toward the polar region,
the null surface shape becomes prolate (or oblate).


The field angular frequency $\Omega_{\rm F}$ is, indeed,
deeply related to the accretion conditions.
For instance, 
in order to get adequate jet efficiency from the magnetic field 
in the funnel above the black hole for a radio loud 
active galactic nuclei (AGN), 
one needs significant flux compression from the lateral boundary imposed 
by the accretion flow and corona \citep{mckinney12,sikora13}.
These lateral boundary conditions and the plasma 
injection process will modify $\Omega_{\rm F}$ in the funnel to
$-0.10 \omega_{\rm H}< \Omega_{\rm F} < 0.35\omega_{\rm H}$ 
\citep{mckinney12,bes13}.
For illustrative purposes, we simplify this behavior by choosing 
a constant $\Omega_{\rm F} = 0.3 \omega_{\rm H}$ for $A_\varphi$. 
We acknowledge that this is not a self consistent solution, 
but we believe that it is sufficient 
for demonstrating the physical process that we propose.
In \S~\ref{sec:IC310_omgF}, we consider the cases of 
different $\Omega_{\rm F}$ and compare the results.


\subsection{The stagnation surface}
\label{sec:stagnation}
In addition to the null surface discussed above,
a stagnation surface is also considered to be a plausible place
for particle accelerators to arise.
Since both inflows and outflows start from the stagnation surface
in magnetohydrodynamics (MHD), 
we can expect that the plasma density becomes low, 
thereby leading to a non-vanishing 
$E_\parallel$ there. 
\citet[LR11]{levi11} and \citet[BT15]{brod15} examined this possibility.
In this subsection, we examine only the position of the
stagnation surface,
leaving its electrodynamical plausibility
as a particle accelerator site to \S~\ref{sec:electrodynamics}.

\begin{figure}
 \epsscale{1.0}
 \includegraphics[angle=0,scale=0.43]{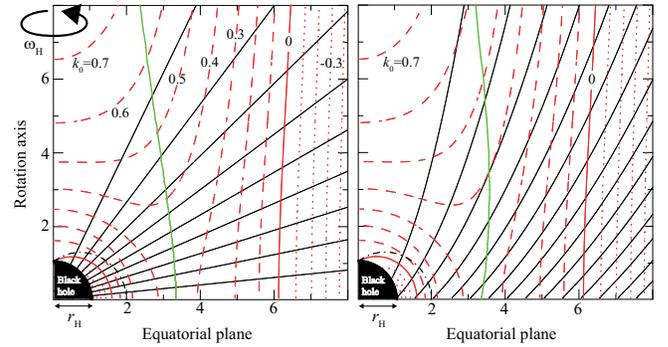}
\caption{
Distribution of the stagnation surface (thick green solid curve)
on the poloidal plane.
Similar figure as figure~\ref{fig:side_null}, 
but equi-$k_0$ potential curves are depicted
with the dashed curves (for positive values)
and dotted ones (for negative values). 
The thick, red solid curves denote the outer and inner 
light surfaces where $k_0$ vanishes.
Stationary MHD flows start from the maximum-$k_0$ point
along the individual magnetic field lines
\citep{taka90}.
The left panel represents the case of a split-monopole 
magnetic field, while the right one a parabolic field. 
\label{fig:side_k0}
}
\end{figure}

In a stationary and axisymmetric black hole magnetosphere,
MHD inflows and outflows start
from the two-dimensional surface on which
\begin{equation}
  k_0{}'=0
  \label{eq:k0prime}
\end{equation}
holds \citep{taka90}, where 
\begin{equation}
  k_0 \equiv -g_{tt}-2g_{t\varphi}\Omega_{\rm F}
  -g_{\varphi\varphi} \Omega_{\rm F}{}^2,
  \label{eq:k0}
\end{equation}
and the prime denotes
the derivative along the poloidal magnetic field line.
The condition~(\ref{eq:k0prime}) is equivalent with imposing
a balance among the gravitational, centrifugal, and Lorentz forces
on the poloidal plane.
In figure~\ref{fig:side_k0},
we plot the contours of $k_0$ for split-monopole and parabolic
poloidal magnetic field lines.
Far away from the horizon, we obtain
$k_0 \approx c^2 - \varpi^2 \Omega_{\rm F}{}^2$,
where $\varpi$ denotes the distance from the rotation axis.
Thus, at large $\varpi$, the potential $k_0$ becomes negative,
as denoted by the red dotted contours
on the right part of each panel.
Note that $k_0$ vanishes
at $\varpi \approx c/\Omega_{\rm F}$.
This two-dimensional surface is called the outer light surface
and depicted by the thick red solid curve in figure~\ref{fig:side_k0}.
If the general-relativistic effects
(i.e., redshift and frame-dragging effects) were negligible,
and if $\Omega_{\rm F}$ is constant for the
magnetic flux function $A_\varphi$,
the outer light surface would distribute cylindrically,
as in pulsar magnetospheres.
Outside the outer light surface, plasma particles must
flow outwards, if they are frozen-in to the magnetic field.
Inside the outer light surface,
$k_0$ becomes positive, as denoted by the dashed contours.
Further inside, close to the horizon,
$k_0$ begins to decrease again due to the strong gravity of the hole.
As a result, another light surface, 
where $k_0$ vanishes, appears, 
as depicted by the thick red solid curve near the horizon.
Inside this inner light surface, 
plasma particles must flow inwards
if they are frozen-in to the magnetic field.

As demonstrated by \citet{taka90},
a stationary and axisymmetric MHD flow flows from a greater
$k_0$ region to a smaller $k_0$ one.
It follows that both the inflows and outflows start from the 
two-dimensional surface on which 
$k_0$ maximizes along the poloidal magnetic field line.
Thus, putting $k_0{}'=0$, we obtain the position of the
stagnation surface.
If figure~\ref{fig:side_k0},
we plot the stagnation surface as the thick green solid curve.
It is clear that the stagnation surface is located at
$r < 5 r_{\rm g}$ in the lower latitudes but at
$r > 5 r_{\rm g}$ in the higher latitudes,
irrespective of the magnetic field line configuration,
as long as $\Omega_{\rm F}$ is a good fraction of $\omega_{\rm H}$,
e.g., $0.25 \omega_{\rm H} < \Omega < 0.50 \omega_{\rm H}$.
It is, however, noteworthy that the numerical simulations
\citep{mckinney12}
and a supporting analytical work \citep{bes13}
claimed $-0.10 \omega_{\rm H} < \Omega < 0.35 \omega_{\rm H}$
in the funnel to represent radio-loud AGN.
As the poloidal field geometry changes from radial
to parabolic, 
the stagnation surface moderately moves away from the rotation axis
in the higher latitudes.

This analytical result was confirmed by 
general relativistic (GR) MHD simulations
\citep{mckinney06,brod15}.
In these numerical works, 
the stagnation surface is time-dependent but stably located
at 5--10 $r_{\rm g}$ with a prolate shape,
as depicted in figure~\ref{fig:side_k0}. 
In addition, 
the position of the stagnation surface changes
as a function of the accretion rate and other numerical settings
such as the initial and boundary conditions.
However, it should be emphasized that these changes are caused
merely through the poloidal magnetic field configuration
and $\Omega_{\rm F}=\Omega_{\rm F}(A_\varphi)$,
the latter of which defines the potential $k_0$.

To examine the Poisson equation~(\ref{eq:pois}),
it is worth noting that the distribution of $\rho_{\rm GJ}$ 
has relatively small dependence on the magnetic field geometry
near the horizon,
because its distribution is essentially governed by the
space-time dragging effect,
as indicated by the red solid, dashed, and dotted contours
in figure~\ref{fig:side_null}
for split-monopole and parabolic fields.
As a result, the gap solution little depends on 
the magnetic field line configuration,
particularly in the higher latitudes,
forming a striking contrast to pulsar outer-gap models,
in which the field line configuration defines $\rho_{\rm GJ}$.

As for $k_0=k_0(r,\theta)$, 
it has no dependence on the field geometry,
as long as $\Omega_{\rm F}$ is constant for all the field lines.
In this particular case, the stagnation surface distributes
more or less similarly for split-monopole and parabolic fields,
as figure~\ref{fig:side_k0} indicates.
Thus, in what follows, we assume a split-monopole field 
and adopt a constant $\Omega_{\rm F}=\Omega_{\rm F}(A_\varphi)$ 
for simplicity, as described in the left panels of 
figures~\ref{fig:side_null} and \ref{fig:side_k0}.
However, it is noteworthy that the lateral boundary conditions 
and the plasma injection in the tenuous funnel alter both the 
magnetic-field distribution near the event horizon and $\Omega_{\rm F}$ 
in an actual solution 
\citep{phinney83,mckinney12,bes13}. 


\subsection{Electrodynamical consideration of gap position}
\label{sec:electrodynamics}
Let us leave the geometrical argument of the stagnation surface
and turn to the electrodynamics of a gap,
which may be formed around the null surface or the stagnation surface.
For this sake, we must consider the Poisson equation~(\ref{eq:pois})
for the non-corotational potential $\Psi$.

If the full gap width $w$ along the magnetic field line
is short compared to the meridional and azimuthal dimensions of the gap,
only the derivatives along the poloidal magnetic field line remains in
equation~(\ref{eq:pois}).
Noting that the gravitational forces are negligible compared
to the electrodynamical forces in the gap
(except for the intimate vicinity of the horizon), 
we can reduce equation~(\ref{eq:pois}) into 
\begin{equation}
  \frac{d E_\parallel}{ds}= 4\pi(\rho-\rho_{\rm GJ}),
  \label{eq:pois_2}
\end{equation}
where $s$ denotes the distance along the poloidal magnetic field line
measured outward in the Boyer-Lindquist coordinates.
We define $s=0$ to indicate the null surface, $r=r_0$.
Thus, the position $s$ that satisfies $r_0+s=r_{\rm H}$ 
represents the event horizon,
because we adopt radial magnetic field lines on the poloidal plane. 
Both the inner boundary (at $s=s_1$) and the outer boundary 
(at $s=s_2$) are free boundaries
and are determined as a function of the accretion rate
by the procedure to be described in \S~\ref{sec:closure}.
If the real charge density $\rho$ is greater (or smaller)
than $\rho_{\rm GJ}$, $E_\parallel$ increases (or decreases)
outwards.

The created $e^\pm$ pairs are separated by $E_\parallel$
in the gap.
Without loss of any generality, we can assume
$ E_\parallel < 0$ in the gap, which is possible if 
$\mbox{\boldmath$B$} \cdot \mbox{\boldmath$L$} > 0$ 
(or exactly speaking, if $A_{\varphi,\theta}>0$)
in the northern hemisphere near the horizon,
where $\mbox{\boldmath$L$}$ denotes the BH's angular momentum vector.
Note that the sign of $E_\parallel$ changes from
pulsar outer gaps,
because the null surface is formed by the frame dragging
instead of the convex topology of the poloidal magnetic field lines.
The negative $E_\parallel$ results in an outwardly decreasing
dimensionless charge density, $\rho/(\Omega B/2\pi c)$.
Since the total current per magnetic flux tube conserves,
the sum of positronic and electronic charge densities
per magnetic flux tube also conserves.
Thus, pair production per unit length,
$d(\rho/B)/ds$, becomes more or less constant for $s$, 
provided that the background soft photon density is roughly homogeneous,
as demonstrated in figure~5 of \citet{hiro99}.
We therefore assume that 
$\rho/B$ decreases linearly with $s$ in the present paper.
In general, to analyze the spatial dependence of $\rho/B$, 
we must solve the Boltzmann equations of electrons, positrons, 
and photons simultaneously 
together with the Poisson equation~(\ref{eq:pois_2}),
as \citet{hiro99} did for pulsar magnetospheres.

In the present paper, we employ a simplified model
such that $\rho/B$ decreases with increasing $s$ linearly as
\begin{equation}
  \frac{\rho(s)}{B(s)}
  =  j \frac{\rho_{\rm GJ}(s_2)}{B(s_2)}
       \frac{2s-s_1-s_2}{s_2-s_1},
  \label{eq:rho_linear}  
\end{equation}
where $0 < j \le 1$.
Note that $s=s_2$ denotes the outer boundary 
and that $\rho(s_2)$ is negative.
If we put $j=0$, we obtain a vacuum gap.
In a non-vacuum gap, $j \ne 0$,
$E_\parallel$ is partially screened by the created and polarized pairs.
If $j=1$, we obtain $\rho(s_2) = \rho_{\rm GJ}(s_2)$.
A super-GJ current, $j>1$, is prohibited,
as will be discussed in \S~\ref{sec:comparison_BT}.

\subsubsection{Vacuum gap around the null surface}
\label{sec:vac_null}
Let us begin with the vacuum case, $j=0$, which leads to $\rho=0$. 
The Poisson equation~(\ref{eq:pois_2}) shows $dE_\parallel/ds>0$ 
(or $dE_\parallel/ds<0$) holds in the outer (or inner) part of the gap,
because $\rho-\rho_{\rm GJ}=-\rho_{\rm GJ}$ becomes positive (or negative)
in the outer (or inner) part of the gap.
Thus, we obtain $E_\parallel<0$ in the gap,
and find that $\vert E_\parallel \vert$ maximizes 
at the point where $\rho$ matches $\rho_{\rm GJ}$,
which is realized at the null surface, $s=0$, in a vacuum gap.
Outside the gap, on the other hand, to meet the force-free condition,
$\rho$ should match $\rho_{\rm GJ}$.
It results in a jump of $dE_\parallel/ds$ at the boundaries, 
which should be compensated by the surface charge there.

\begin{figure}
 \epsscale{1.0}
 \includegraphics[angle=0,scale=0.50]{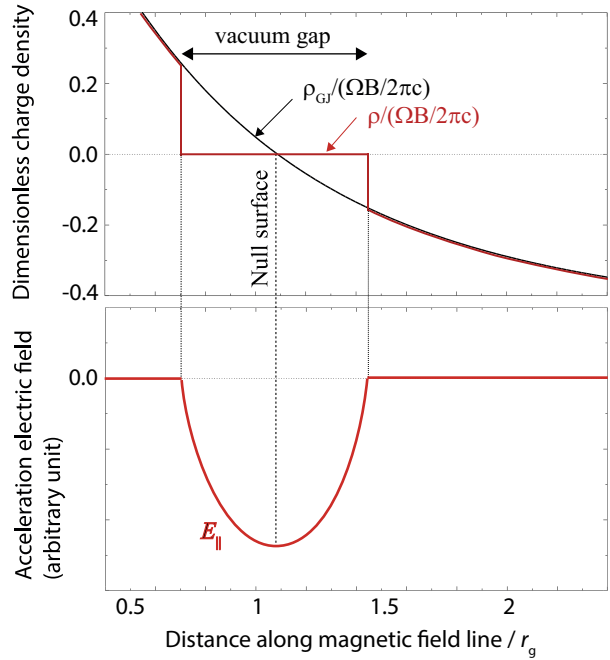}
\caption{
Distribution of the charge density and the magnetic-field-aligned
electric field along the magnetic field. 
A vacuum gap is assumed.
In the upper panel, the black solid curve represents the
Goldreich-Julian charge density per magnetic flux
along the split-monopole magnetic field line at $\theta=15^\circ$,
while the red solid curve does the dimensionless real charge density
per magnetic flux.
In the lower panel, the resultant $E_\parallel$ is illustrated
so that the distribution of the $\rho-\rho_{\rm GJ}$ sign
may be reflected.
Since the gap closure condition (\S~\ref{sec:closure})
cannot be satisfied for $\rho=0$, 
we cannot determine the full gap width $w$.
Thus, we can only illustrate the $E_\parallel$ distribution
by hand for a vacuum gap.
\label{fig:rho_1}
}
\end{figure}

For analytical purpose, in this particular subsection 
(\S~\ref{sec:vac_null}), 
we assume that the full gap width, $w$,
is much small compared to $r_{\rm g}$, and
expand equation~(\ref{eq:pois_2}) around the null surface ($s=0$)
to obtain
\begin{equation}
  \frac{d E_\parallel}{ds}= -4\pi \rho_{\rm GJ}{}' s,
  \label{eq:pois_3}
\end{equation}
where $\rho_{\rm GJ}{}'$ denotes the derivative of $\rho_{\rm GJ}$
at $s=0$.
Imposing $E_\parallel=0$ at both boundaries ($s=\pm w/2$), we obtain
\begin{equation}
  E_\parallel(s)
  = 2\pi \rho_{\rm GJ}{}' 
    \left[ (w/2)^2-s^2 \right] .
  \label{eq:Epara}
\end{equation}
The distribution of $\rho/B$, $\rho_{\rm GJ}/B$, and $E_\parallel$
are illustrated in figure~\ref{fig:rho_1}.
As the gap width $w$ increases,
$\vert E_\parallel \vert$ increases quadratically.
Integrating equation~(\ref{eq:Epara}) over $s$ from 
$s_1=-w/2$ to $s_2=+w/2$,
we obtain the electric potential drop in the gap,
\begin{equation}
  V_{\rm gap}
  = \frac83 \pi \rho_{\rm GJ}{}' (w/2)^3
  \approx \left( \frac{w/2}{r_{\rm H}} \right)^3 
          (\mbox{EMF}), 
  \label{eq:Vgap}
\end{equation}
where the electromotive force exerted on the hemisphere of the horizon
is given by
\begin{equation}
  \mbox{EMF} \equiv \frac{\Omega_{\rm F}}{c} r_{\rm H}{}^2 B,
\end{equation}
and 
 \begin{equation}
   \rho_{\rm GJ}{}' 
   \approx \frac{\Omega_{\rm F}B}{2\pi c}
           \frac{1}{r_{\rm H}}
 \end{equation}
is used.
Although we expand $\rho_{\rm GJ}$ around $s=0$,
the argument on $V_{\rm gap}$ is basically valid
even when $w > r_{\rm g}$,
because 
$\rho_{\rm GJ}$ changes from $0$ (at the null surface) to 
$\approx (\omega_{\rm H}-\Omega_{\rm F})B/(2 \pi c)$ 
(at the horizon) in any case.

If we assume some $w$, we can solve $E_\parallel(s)$ 
with equation~(\ref{eq:pois_2}).
For a vacuum gap, there is no photon flux that is needed to realize
the pair creation cascade that closes the gap.
Thus, we cannot solve $w$ for a given accretion rate.
Thus, we can only illustrate by hand how $\rho(s)$ and $E_\parallel(s)$
are related in figure~\ref{fig:rho_1}.
A vacuum gap can be naturally formed around the null surface, 
because $dE_\parallel/ds$ changes sign at the null surface. 
There remains, however, another problem how to supply 
the charges that emit photons
and how to realize a force-free magnetosphere outside the gap.
This motivates us to consider pair creation inside and outside the gap.

\subsubsection{Non-vacuum gap around the null surface}
\label{sec:nonvac_null}
To pursue the supply of charges into the magnetosphere,
we next consider a non-vacuum gap, $0 < j \le 1$.
In a non-vacuum gap, $\rho_{\rm GJ}$ is partly canceled by $\rho$, 
leading to a smaller $\vert E_\parallel \vert$ than the
vacuum value.
To grasp the relationship between $\rho(s)$ and $E_\parallel(s)$,
we only show the representative results of their distribution 
in this section, 
leaving the further details of electrodynamics
in \S~\ref{sec:app_IC310}.

\begin{figure}
 \plotone{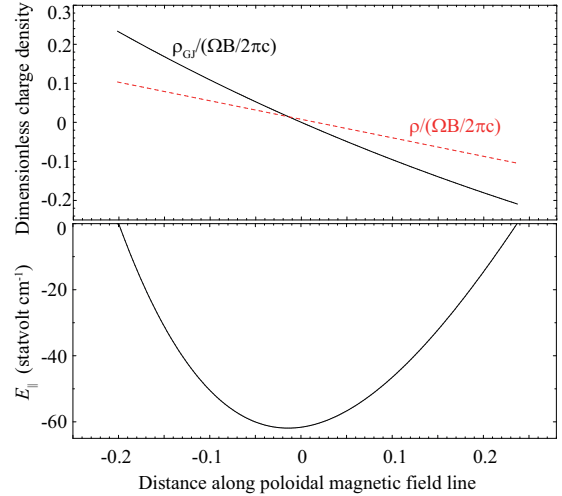}
\caption{
Charge densities and acceleration electric field
in a non-vacuum gap with mild screening, $j=0.5$.
Both the inner and outer boundaries are solved as a free boundary problem
for the dimensionless accretion rate, $\dot{m}=3.16 \times 10^{-5}$
(for the detailed methods, see \S~\ref{sec:app_IC310}).
{\it Top panel:} 
Charge density per magnetic flux tube (red dashed curve),
and the Goldreich-Julian charge density per magnetic flux tube
(black solid curve)
as a function of the distance $s$ from the null surface.
{\it Bottom panel:} Magnetic-field-aligned electric field.
\label{fig:rho_2}
}
\end{figure}

We solve the gap by the method described in \S~\ref{sec:app_IC310}
for the BH magnetosphere of IC~310,
adopting $j=0.5$ and and dimensionless accretion rate
$\dot{m}=3.16 \times 10^{-5}$,
where
\begin{equation}
  \dot{m} \equiv \frac{\dot{M}}{\dot{M}_{\rm Edd}};
  \label{eq:mdot}
\end{equation}
$\dot{M}$ denotes the mass accretion rate, 
$\dot{M}_{\rm Edd}$ the Eddington accretion rate
\begin{equation}
  \dot{M}_{\rm Edd} \equiv \frac{L_{\rm Edd}}{\eta_{\rm eff}c^2}
  = 1.39 \times 10^{27} M_9 \mbox{g s}^{-1},
  \label{eq:mdot_edd}
\end{equation}
$L_{\rm Edd}$ the Eddington luminosity, 
and the conversion efficiency is assumed to be 
$\eta_{\rm eff}=0.1$.
The resultant $\rho(s)$ and $E_\parallel(s)$ distribution
is presented in figure~\ref{fig:rho_2},
The gap inner and outer boundaries are located at
$s_1=-0.201 r_{\rm g}$ and $s_2= 2.345 r_{\rm g}$, respectively.
Quantities $\rho/B$, $\rho_{\rm GJ}/B$, and $E_\parallel$
are depicted only within the gap, $s_1 < s < s_2$.
The maximum $E_\parallel$ becomes 
$6.19 \times 10^1 \mbox{ statvolt cm}^{-1}$.
Because of the screening by the discharged pairs,
this value is about half of the vacuum-case (i.e., $j=0$ case) value
$E_\parallel(0)=1.35 \times 10^2 \mbox{ statvolt cm}^{-1}$ 
for the same $w/2=s_2$ (eq.~[\ref{eq:Epara}]).
The potential drop becomes $V_{\rm gap}= 4.34 \times 10^{14}$~statvolt
for the $j=0.5$ case.

\begin{figure}
 \plotone{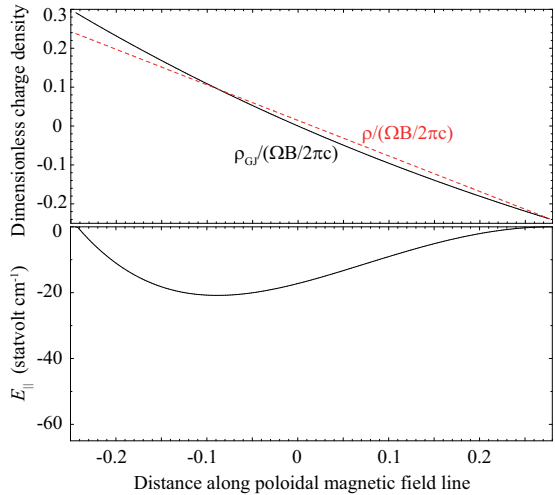}
\caption{
Similar figure as figure~\ref{fig:rho_2}
but with the strongest screening, $j=1.0$.
Other parameters than $j$ are the same as figure~\ref{fig:rho_2}.
Because of less efficient gamma-ray emission, the gap width enlarges
compared to the $j=0.5$ case (fig.~\ref{fig:rho_2}).
\label{fig:rho_3}
}
\end{figure}

We also present the case of the marginally super-GJ case, $j=1$, 
in figure~\ref{fig:rho_3}.
Since $\rho=\rho_{\rm GJ}$ holes at the outer boundary,
$dE_\parallel/ds$ vanishes there,
requiring no surface charge.
The maximum $E_\parallel$ becomes 
$2.08 \times 10^1 \mbox{ statvolt cm}^{-1}$,
and the potential drop becomes $V_{\rm gap}= 1.72 \times 10^{14}$~V.
Note that the total luminosity of the gap is
given by the product of $V_{\rm gap}$ 
and the current flowing through the gap,
the latter of which is proportional to $j$.
Since the product $j V_{\rm gap}$ decreases only 20~\%
from $j=0.5$ to $j=1$, 
we find that the gap luminosity depends on $j$ relatively weakly,
as long as $0.5 \le j \le 1$.
We thus adopt $j=1$ as the representative value
in the present paper.

Figures~\ref{fig:rho_2} and \ref{fig:rho_3}
show that the gap width $w=s_2-s_1$ slightly increases
from $j=0.5$ to $j=1.0$.
This is because the weaker $E_\parallel$ for $j=1$ 
results in less energetic IC photons,
which require longer mean-free path to materialize as pairs,
thereby increasing $w$.
However, at the same time, the doubled particle flux of 
the $j=1$ case (from the $j=0.5$ case)
results in an increased IC photon flux,
which in turn contribute to decrease $w$
(for details, see \citet{hiro13}).
Because of such negative feedback effects,
gap solution exists in a vast expanse of parameter space
in BH and pulsar magnetospheres.
For this reason, classic pulsar vacuum outer-gap model
works qualitatively well, because the errors incurred
by the vacuum assumption and the GJ charge density in the gap
partly cancel each other and keep the conclusions
more or less close to the non-vacuum gap model.

%

The maximum current density flowing through the gap,
is limited by
$c \vert \rho_{\rm GJ}(s=s_2) \vert$.
Thus, the wider the gap extends, the greater 
$c \vert \rho_{\rm GJ}(s_2) \vert$ becomes.
For example, near the null surface, 
we can expand $\rho_{\rm GJ}$ to obtain
\begin{equation}
  c \vert \rho_{\rm GJ}(s_2) \vert
  \approx c \vert \rho_{\rm GJ}{}' \vert s_2.
  \label{eq:jmax}
\end{equation} 
Integrating over the hemisphere of the horizon, 
we obtain the total current flowing in the entire magnetosphere,
\begin{equation}
  J_{\rm max} 
  \approx (c \vert \rho_{\rm GJ}{}' \vert s_2) \cdot 2\pi r_{\rm H}{}^2
  \approx J_{\rm tot} \frac{s_2}{r_{\rm H}},
  \label{eq:Jmax}
\end{equation} 
where 
\begin{equation}
  J_{\rm tot} 
  \equiv \frac{(\omega_{\rm H}-\Omega_{\rm F})B}{2\pi}
         \cdot 2\pi r_{\rm H}{}^2
  = (\omega_{\rm H}-\Omega_{\rm F}) B r_{\rm H}{}^2
\end{equation}
represent the maximally possible electric current flowing along 
the magnetic field lines threading the horizon.
The $\omega_{\rm H}-\Omega_{\rm F}$ factor in $J_{\rm tot}$
comes from the fact that $\rho_{\rm GJ}$ is proportional to
$\omega_{\rm H}-\Omega_{\rm F}$ at the horizon.

Combining equations~(\ref{eq:Vgap}) and (\ref{eq:Jmax}),
we obtain the maximally possible luminosity of the gap,
\begin{equation}
  L_{\rm gap, max}
  = V_{\rm gap} J_{\rm max} 
  \approx
    (\mbox{EMF}) \cdot J_{\rm tot} 
    \left( \frac{s_2}{r_{\rm H}} \right)^4
  = L_{\rm sd} \left( \frac{s_2}{r_{\rm H}} \right)^4.
  \label{eq:Lgap}
\end{equation}
If $E_\parallel$ took the vacuum value (i.e., eq.~[\ref{eq:Epara}])
but the current was $J_{\rm max}$,
which would contradict each other,
the gap luminosity would become $L_{\rm gap, max}$.
In a non-vacuum gap, $E_\parallel$ is less than the vacuum value
and the current is less than $J_{\rm max}$;
thus, the gap luminosity should be less than equation~(\ref{eq:Lgap}).
The $s_2{}^4 \approx (w/2)^4$ dependence comes from the fact 
that $E_\parallel$ is proportional to $s_2{}^2$, 
that the potential drop is proportional to $E_\parallel s_2$, 
and that the maximum current is also proportional to $s_2$
(eq.~[\ref{eq:Jmax}]). 

In short, a non-vacuum gap can be formed 
around the null surface, because $\rho-\rho_{\rm GJ}$,
and hence $dE_\parallel/ds$ 
naturally changes sign within the gap,
as demonstrated in figures~\ref{fig:rho_2} and \ref{fig:rho_3}.

\subsubsection{Gap formation away from the null surface}
\label{sec:nonvac_stag}
\begin{figure}
 \epsscale{1.0}
 \includegraphics[angle=0,scale=0.50]{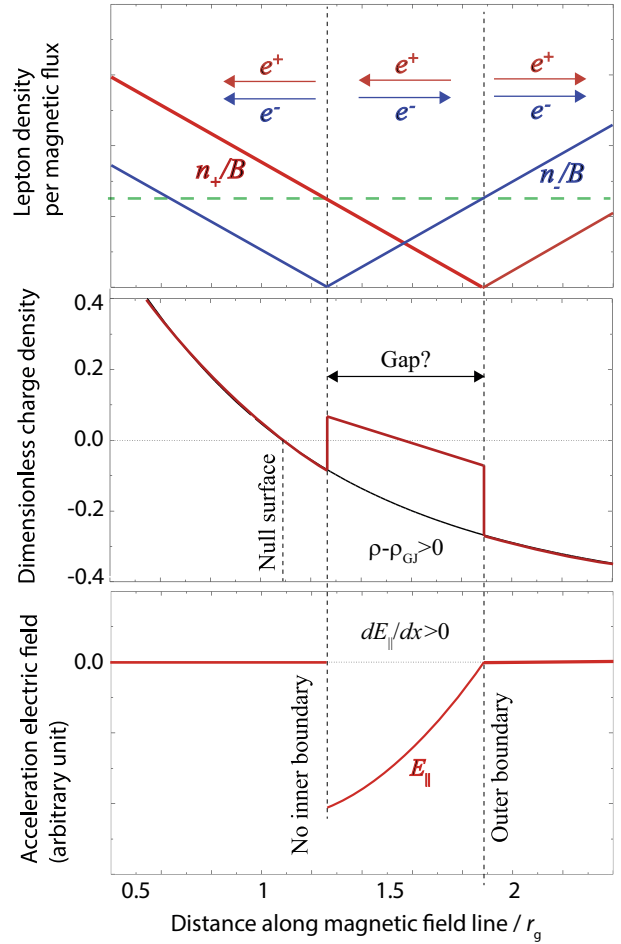}
\caption{
Similar figure as figure~\ref{fig:rho_2},
but the gap position is artificially shifted from the null surface.
In the shifted gap, a single-signed magnetic-field-aligned
electric field, $E_\parallel$,
accelerates electrons and positrons in the opposite directions.
\label{fig:rho_4}
}
\end{figure}

Let us next examine the possibility of the formation of a gap
away from the null surface, e.g., around the stagnation surface.
In figure~\ref{fig:rho_4}, we sketch the possibility of 
the formation of a gap around $s \approx 1.57 r_{\rm g}$.
If there is no particle injection across the boundaries,
only electrons exist at the outer boundary,
and only positrons at the inner boundary (top panel).
In this case, $\rho-\rho_{\rm GJ}$ is positive definite
(middle panel),
resulting in a monotonically increasing $E_\parallel$
in the gap (bottom panel).
Thus, the inner boundary cannot be formed in this case
if we set $E_\parallel=0$ at the outer boundary.

\begin{figure}
 \epsscale{1.0}
 \includegraphics[angle=0,scale=0.50]{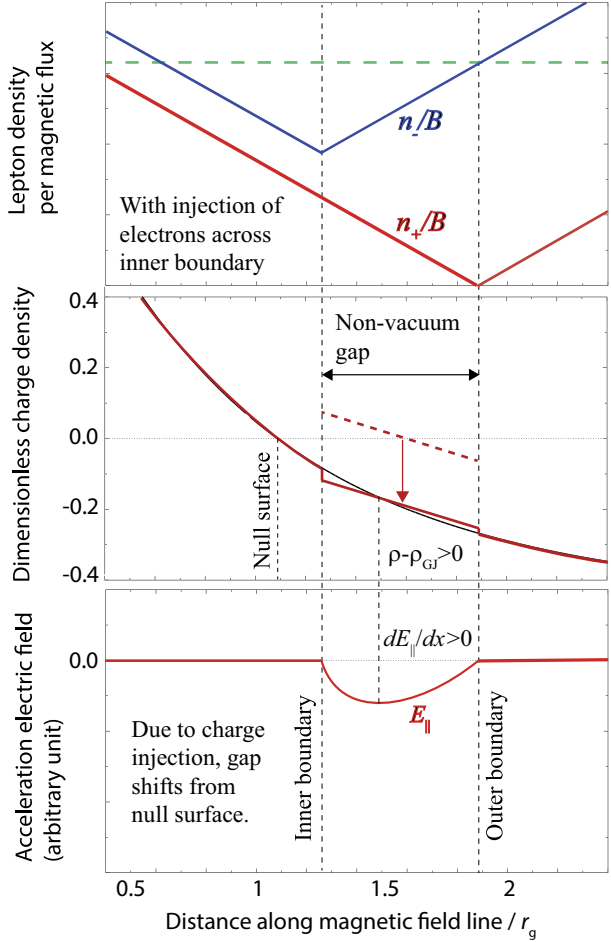}
\caption{
Similar figure as figure~\ref{fig:rho_2},
but an electric current is injected across 
the inner boundary so that $\rho-\rho_{\rm GJ}$ may change sign
within the gap.
\label{fig:rho_5}
}
\end{figure}

To change the sign of $\rho-\rho_{\rm GJ}$,
we must consider the injection of electrons
across the inner boundary,
in the same way as in pulsar outer gap model
\citep{hiro01a}.
It is possible if a small-amplitude, residual $E_\parallel$ 
separate the charges at $s<s_1$, 
and if a fraction of the returned electrons enter the gap.
In this case, the position of the gap center is
determined solely by the injected current across the boundary.
Thus, if the injected current density per magnetic flux tube,
$ce(n_-/B)_1$, coincidentally matches $c\rho_{\rm GJ}/B$
at the stagnation surface,
the gap appears around the stagnation surface,
where the subscript $1$ shows that the quantity is evaluated
at the inner boundary.
Because $\rho-\rho_{\rm GJ}$ changes sign 
(middle panel of fig.~\ref{fig:rho_5}), 
$E_\parallel$ also changes sign (bottom panel)
to close the gap.
We interpret that the treatment of LR11 and BT15 
corresponds to this case, 
although it is not explicitly mentioned in their papers.

In short, the position of a gap shifts outwards (or inwards)
if there is a leptonic current injection across the
inner (or outer) boundary.
The center of the gap is determined by the strength of the
injected current.
Only when the injected current density $ce(n_-/B)$ at the inner boundary
matches $c\rho_{\rm GJ}/B$ at the stagnation surface,
the gap center can be located at the stagnation surface.

%
In BH gap models, there have been two main ideas about the position
of the gap.
One considers gaps at the null surface
\citep[HO98]{bes92},
and the other at the separation surface (LR11,BT15).
In this paper, we pursue the first possibility,
because there is no essential difference in gap electrodynamics
between the two cases, and because no current injection
may be the simplest assumption.
For instance, the gap closure condition (\S~\ref{sec:closure}) 
is much simplified in the former case,
because we do not have to consider additional $\gamma$-ray emission
by the injected charges.


\section{Radiatively inefficient accretion}
\label{sec:RIAF}
In this paper, we consider a situation in which plasma accretion
takes place in the lower latitudes, that is, near the
rotational equator with a certain vertical thickness
\citep{krolik05,mckinney06,mckinney07a,mckinney07b,
punsly09,punsly11}.
Such an accretion cannot penetrate into the higher latitudes,
that is, in the polar funnel,
because the centrifugal-force barrier prevents plasma accretion 
towards the rotation axis, and
because the time scale for magnetic Rayleigh-Taylor instability
or turbulent diffusion is long 
compared to the dynamical time scale of accretion.
In this evacuated funnel, the poloidal magnetic field lines
resemble a split monopole in a time-averaged sense
\citep{hirose04,mckinney12}.
The lower-latitude accretion emits MeV photons into the
higher latitudes, supplying electron-positron pairs in the funnel.
If the pair density in the funnel becomes less than the
GJ number density, an electric field arises
along the magnetic field.
Thus, a plasma accretion in the lower altitudes and
a gap formation in the higher altitudes are compatible
in a BH magnetosphere.

From multi-frequency radio observations, 
the jet power of IC~310 is estimated to be
$L_{\rm jet} = 2 \times 10^{42} \mbox{ergs s}^{-1}$
\citep{sijb98}.
If this jet is accretion-powered, 
we find that the dimensionless accretion rate will be
\begin{equation}
  \dot{m} = \eta_{\rm j}{}^{-1}
            L_{\rm jet}/L_{\rm Edd} 
          = 1.6 \times 10^{-4} 
            \left( \frac{\eta_{\rm j}}{0.1} \right)^{-1}
            M_9{}^{-1},
  \label{mdot_upper}
\end{equation}
where $\eta_{\rm j}$ denotes the energy conversion efficiency
from accretion into jet \citep{mckinney12}.
Thus, we can evaluate $\dot{m} \sim 10^{-4}$ or slightly greater,
depending on the value of $\eta_{\rm j}$.

If the jet is BH-rotation-powered, 
the Blandford-Znajek process energizes the jet.
Its power can be estimated to be \citep{bla76}
\begin{equation}
  L_{\rm sd} 
  \approx 10^{42} \left( \frac{a}{M} \right)^2
                  \left( \frac{M_9}{0.3} \right)^2
                  B_3{}^2 \mbox{ergs s}^{-1},
  \label{eq:L_BZ}
\end{equation}
where $B_3=B/(10^3\mbox{ G})$.
If the magnetic pressure is in equilibrium with the gas pressure,
we obtain the equilibrium magnetic field strength (LR11) 
\begin{equation}
  B_{\rm eq}
  \approx 4 \times 10^4 \dot{m}^{1/2} M_9{}^{-1/2} \mbox{ G}.
  \label{eq:B_eq}
\end{equation}
Equating $L_{\rm sd}$ with $L_{\rm jet}$, 
and setting $B \approx B_{\rm eq}$,
we obtain $B_3 \approx 1.4$ and
$\dot{m} \sim 4 \times 10^{-4}$.

Combining these two estimates,
we can infer the accretion rate to be
$\dot{m} \sim 2 \times 10^{-4}$
on {\it time-averaged sense}.
Nevertheless, it is worth noting that this argument 
does not restrict the value of $\dot{m}$ 
on much shorter timescales than the jet propagation times scale
at radio-emitting regions,
the latter of which is much longer than the
light crossing time of the horizon.

At such a low accretion rate, $\dot{m} \sim 10^{-4}$ or less,
the accretion flow becomes radiatively inefficient 
\citep{ichimaru77,narayan94,narayan95,narayan97,
       nakamura97}.
In the present paper, for analytical purpose, we adopt 
the self-similar solution of an advection-dominated accretion flow
(ADAF) in Newtonian approximation \citep{mahad97},
skipping to incorporate general relativistic corrections
\citep{manmoto00,li08}
into the ADAF model.
Such plasma accretion takes place in the lower latitudes
(i.e., near the rotational equator with a certain vertical thickness)
and emit photons into many directions including the higher latitudes
(i.e., the polar funnel) where jets are launched.
Thus, to estimate the density of the created pairs in the higher latitudes 
(e.g., around the colatitude $\theta \sim 10^\circ$ or $20^\circ$), 
we must examine the flux of the MeV photons 
emitted by the ADAF.

To evaluate the flux of ADAF MeV photons,
we consider the free-free emission in the inner region of the ADAF. 
The free-free emission spectrum peaks around
$\nu_{\rm p} = k T_{\rm e}/h$,
where $T_{\rm e}$ denotes the electron temperature,
$k$ the Boltzmann constant, and $h$ the Planck constant.
Near the peak, eq.~(30) of \citet{mahad97} gives 
the soft photon luminosity
\begin{equation}
  L_{\rm s}= 2.06 \times 10^{38} T_{{\rm e},10} m \dot{m}^2
            \mbox{ergs s}^{-1}
  \label{eq:Ls}
\end{equation}
if the viscosity parameter is $\alpha=0.3$,
$c_1= 0.5$, $r_{\rm max}/r_{\rm min}=10^3/3$, and 
$T_{{\rm e},10} \equiv T_{\rm e}/(10^{10}~K)>0.5$
(see \citet{mahad97} for details).
Thus, around $r=6r_{\rm g}$,
the MeV photon number density becomes
\begin{equation}
  n_{\rm s}= 4.72 \times 10^{11} M_9{}^{-1} \dot{m}{}^2
            \left( \frac{r}{6 r_{\rm g}} \right)^{-2} \mbox{cm}^{-1}.
  \label{eq:ns}
\end{equation}
Following the logic around equation~(6) of LR11 
we obtain the number density of pairs created by the
collisions of these MeV photons,
\begin{equation}
  n_\pm= 8.88 \times 10^{12} M_9{}^{-1} \dot{m}{}^4
            \left( \frac{r}{6 r_{\rm g}} \right)^{-3} 
        \mbox{cm}^{-1}.
  \label{eq:npair}
\end{equation}

Evaluating $B$ with equation~(\ref{eq:B_eq}),
we obtain the GJ number density, 
\begin{equation}
  n_{\rm GJ}
  = 4.41 \times 10^{-2} M_9{}^{-3/2} \dot{m}{}^{1/2}
                \frac{a}{M} \frac{\Omega_{\rm F}}{\omega_{\rm H}}
        \mbox{cm}^{-1},
  \label{eq:nGJ}
\end{equation}
Thus, we obtain
\begin{equation}
  \frac{n_\pm}{n_{\rm GJ}}
  = 2.01 \times 10^{14} M_9{}^{1/2} \dot{m}{}^{7/2}
                \left( \frac{a}{M} \right)^{-1}
                \left( \frac{\Omega_{\rm F}}{\omega_{\rm H}} \right)^{-1}
                \left( \frac{r}{6 r_{\rm g}} \right)^{-3} . 
  \label{eq:n_ratio}
\end{equation}
It follows that $n_\pm < n_{\rm GJ}$ approximately holds 
if $\dot{m} < 10^{-4}$ for $M_9=0.3$.
In another word, the {\it time-averaged} 
value of $\dot{m} \sim 2 \times 10^{-4}$
gives an order of magnitude greater pair density
than the GJ value.
Therefore, for IC~310,
we expect that the BH gap is quenched by the ADAF-provided pairs 
along any magnetic field lines
during a substantial portion of time.

It is, however, noteworthy that 
$n_\pm/n_{\rm GJ}$ strongly depends on $\dot{m}$.
For instance, if $\dot{m}$ is halved,
$n_\pm/n_{\rm GJ}$ decreases one order of magnitude.
Thus, it is worth considering the cases in which
a smaller accretion rate, $\dot{m} \ll 10^{-4}$,
is achieved in a limited space and time.
For instance, because a lower-latitude accretion 
in a geometrically thick disk is known to be
highly variable due to turbulence
\citep{hirose04,mckinney07a}, 
$\dot{m} \ll 10^{-4}$ may be achieved 
during a short period of time.
What is more,
it is suggested by three-dimensional MHD simulations
that an accretion of magnetized plasmas tends to be
highly variable near the horizon
particularly when the BH spin approaches the 
extreme value, $a \approx 0.998 r_{\rm g}$
\citep{krolik05}.
It was also analytically pointed out
that a strongly magnetized plasma accretion tends to be 
highly variable near the horizon \citep{hiro93}.
This is because the meridional current suffers a large-amplitude 
fluctuation (compared to e.g., the radial magnetic field fluctuations)
at the fast-magnetosonic surface,
which is located in the vicinity of the horizon
in a magnetically dominated magnetosphere \citep{phinney83}.
This results in a strong perturbation of the Lorentz force,
leading to a large radial acceleration of the MHD inflow. 
This phenomena is solely general relativistic (GR) in the sense
that the plasma inertia and the causality at the horizon are essential.
Thus, it will not happen in non-GR trans-fast-magnetosonic flows.

Since our viewing angle of IC~310 is constrained to be
between $10^\circ$ and $20^\circ$ with respect to the
jet axis \citep{alek14b},
we assume that $\dot{m} < 10^{-4}$ is achieved 
in the higher latitudes intermittently.
In this case, the BH gap of IC~310 will be switched on
with a small duty cycle 
(i.e., during a small fraction of time).

\section{Gap electrodynamics}
\label{sec:phys}
We formulate a stationary particle accelerator exerted around
a rotating BH in this section.

\subsection{Saturated Lorentz factors}
\label{sec:Lf}
If the curvature drag force balances $e E_\parallel$,
the electron Lorentz factors are saturated at
the curvature-limited value,
\begin{equation}
 \gamma_{\rm curv} 
   \equiv 
   \left( \frac{3 \rho_{\rm c}^2}{2e} E_\parallel 
   \right)^{1/4},
  \label{eq:Lcv}
\end{equation}
where $\rho_{\rm c}$ denotes the curvature radius of the 
three-dimensional particle path,
which becomes comparable to the curvature radius of the 
magnetic field lines in a rotating magnetosphere \citep{hiro11}.
Although the split-monopole solution specifies 
not only the poloidal magnetic field components 
but also the toroidal (i.e., azimuthal) one,
we set $\rho_{\rm c}$ as a free parameter in the present paper,
because a time-dependence accretion onto a BH will 
have a spatially and temporary varying magnetic field near the horizon
\citep{krolik05}.

If the IC drag force balances $e E_\parallel$, 
on the other hand,
the saturated Lorentz factor, $\gamma_{\rm IC}$, 
should be implicitly solved from the balance
\begin{equation}
   \frac{1}{c}
   \int_{E_{\rm min}}^{\gamma_{\rm IC} m_{\rm e} c^2}
      E_\gamma dE_\gamma
    \int_{E_{\rm min}}^{E_{\rm max}}
      \frac{d\sigma}{dE_\gamma}
      \frac{dF_{\rm s}}{dE_{\rm s}} dE_{\rm s}
    = e E_\parallel,
  \label{eq:Lic}
\end{equation}
where $m_{\rm e}$ denotes the rest mass of the electron,
$E_\gamma$ and $E_{\rm s}$ the (emitted) hard-photon and 
(input) soft-photon energies, respectively,
$E_{\rm min}$ and $E_{\rm max}$ the lower and upper bound energies
we are considering,
$d\sigma/dE_\gamma$ the Klein-Nishina differential cross section,
$dF_{\rm s}/dE_{\rm s}$ 
the differential flux of the background soft photon field.
Note that $d\sigma/dE_\gamma$ has $\gamma_{\rm IC}$ dependence.
In \S~\ref{sec:app_IC310},
we demonstrate that the IC scatterings mainly take place in the
extreme Klein-Nishina regime.

We assume that the input ADAF photon field \citep{mahad97}
is isotropic in the BH rest frame.
To cover the photon energies
from radio wavelengths (of ADAF emission) to the VHE,
we adopt $E_{\rm min}=5.11 \times 10^{-6}$~eV
and $E_{\rm max}=5.11 \times 10^{17}$~eV,
dividing the energies logarithmically into 198 bins.
To estimate the normalization of the ADAF photon differential flux 
at the gap,
we divide the total ADAF differential luminosity 
\citep{mahad97} by 
$4 \pi (10 r_{\rm g})^2$,
assuming that the ADAF emission flux is homogeneous
at $r<10r_{\rm g}$ but reduces by 
the $(r/10r_{\rm g})^{-2}$ factor at $r>10r_{\rm g}$;
this assumption is reasonable, 
because the sub-millimeter-IR photons,
which most efficiently contribute for both IC scatterings
and photon-photon pair creation,
are emitted from the inner-most region of ADAF, 
$r \approx 10r_{\rm g}$.
We compute both photon-photon pair creation and IC scatterings, 
from the gap to a large enough 
Boyer-Lindquist radius, $r=60r_{\rm g}$.
If we evaluated the ADAF flux at $r=6r_{\rm g}$
(instead of $10 r_{\rm g}$),
the similar solutions would be obtained at smaller $\dot{m}$.
Nevertheless, emission properties of the gap
(e.g., spectrum near the critical accretion rate)
little changes by the normalization of the ADAF photon flux.
Note that the charge-starvation condition $n_\pm/n_{\rm GJ}<1$ 
(eq.~[\ref{eq:n_ratio}]) 
is more easily satisfied if the ADAF flux is normalized at $r=10 r_{\rm g}$
(than at $r=6r_{\rm g}$).
See also \S~\ref{sec:X-ray} for the impact of the normalization of the
soft photon field on the charge-starvation condition.

The actual Lorentz factor is computed by
\begin{equation}
  \gamma^{-1}= \gamma_{\rm curv}{}^{-1}+\gamma_{\rm IC}{}^{-1},
  \label{eq:Lf}
\end{equation}
which is almost equivalent with
$\gamma= \min(\gamma_{\rm curv},\gamma_{\rm IC})$.

\subsection{Curvature process}
\label{sec:curv}
At a Lorentz factor $\gamma$, a single electron emits 
\citep{rybicki79}
\begin{equation}
  (P_\nu)_{\rm cv}
  = \frac{\sqrt{3} e^2}{h\nu} 
    \frac{\gamma}{\rho_{\rm c}}
    F\left(\frac{\nu}{\nu_{\rm c}}\right)
  \label{eq:dFdnu_cv}
\end{equation}
photons per unit frequency $\nu$ per unit time,
where 
\begin{equation}
  F(x) \equiv x \int_x^\infty K_{5/3}(\xi) d\xi,
  \label{eq:def_F}
\end{equation}
$x \equiv \nu / \nu_{\rm c}$ and $K_{5/3}$ denotes the
modified Bessel function of $5/3$ order.
The characteristic frequency is defined as
\begin{equation}
  \nu_{\rm c}
  \equiv \frac{3}{4\pi}
         \frac{\gamma^3 c}{\rho_{\rm c}} .
  \label{eq:def_nuc}
\end{equation}

Particles have different Lorentz factors at different positions.
We compute the evolution of $\gamma$ as a function of position,
assuming that electrons are created homogeneously at each position.
If pairs are created at $s=s_{\rm cr}$, 
the electrons are accelerated outward 
to attain the Lorentz factor at $s$,
\begin{equation}
  \gamma(s)= \min \left( \frac{-e}{m_{\rm e}c^2} 
                         \int_{s_{\rm cr}}^{s} E_\parallel(s) ds,
                         \gamma_{\rm curv} 
                  \right) . 
  \label{eq:gam_s}
\end{equation}

To count up the photons emitted from various positions,
we divide the full gap width into $n_x$ parts along the particle path.
Accordingly, we can compute the number of photons emitted outwardly
in the $i$-th energy bin per unit time by 
\begin{equation}
  (N_{\rm cv})_i
  = \frac{N_{\rm e}}{n_x} \int_{\nu_{i-1}}^{\nu_i} (P_\nu)_{\rm cv} d\nu ,
  \label{eq:dFcv}
\end{equation}
where $N_{\rm e}$ denotes the total number of electrons 
existing in the entire gap,
$\nu_{i-1}$ and $\nu_i$ the lower and upper bound of the $i$-th
photon energy bin.
In \S~\ref{sec:creation}, we will describe
how $N_{\rm e}$ is computed.

\subsection{Inverse-Compton process}
\label{sec:IC}
Next, we consider the emission properties of the IC process.
At a Lorentz factor $\gamma$, a single electron emits 
\begin{equation}
  (P_\nu)_{\rm IC}
  = \frac{3\sigma_{\rm T}}{16}
    (1-\beta \mu_{\rm c})
    \int_0^\infty dE_{\rm s} \frac{dF_{\rm s}}{dE_{\rm s}}
    \int_{-1}^{1} dx^\ast
    \int_0^{2\pi} d\alpha^\ast f_{\rm IC}
  \label{eq:dFdnu_ic}
\end{equation}
photons per unit frequency per unit time,
where 
$f_{\rm IC}= f_{\rm IC}(x^\ast,\alpha^\ast,E_{\rm s},\mu_{\rm c},\gamma)$
is defined by equations~(26)--(28) of \citet{hiro03},
and $\mu_{\rm c}$ denotes the cosine of the collision angle.
Assuming isotropic IC scatterings in the BH rest frame
when we average over the collision angles,
we obtain the number of photons emitted outwardly in the $i$-th
energy bin per unit time as follows:
\begin{equation}
  (N_{\rm IC})_i
  = \frac{N_{\rm e}}{n_x} 
    \int_{\nu_{i-1}}^{\nu_i}  d\nu 
    \frac12 \int_{-1}^{1} (P_\nu)_{\rm IC} d\mu_{\rm c}.
  \label{eq:dFic}
\end{equation}


\subsection{Photon-photon pair creation}
\label{sec:creation}
The $\gamma$-rays emitted via curvature or IC process,
are partly absorbed by colliding with the ADAF soft photons.
The absorption optical depth for the photons in the $i$-th energy bin
becomes
\begin{equation}
  \tau_i
  = \frac{s_2-s_1}{c} \cdot \frac12
    \int_{-1}^1 d\mu_{\rm c}
    \int_{E_{\rm th}}^\infty \sigma_{\rm p}(E_i,E_{\rm s},\mu_{\rm c})
                         \frac{dF_{\rm s}}{dE_{\rm s}} dE_{\rm s},
  \label{eq:tau}
\end{equation}
where $\sigma_{\rm p}$ refers to 
the photon-photon pair creation mean-free path \citep{jauch55},
$E_i=h(\nu_{i-1}+\nu_i)/2$ denotes the $\gamma$-ray energy.
The threshold energy is given by
$E_{\rm th} = [2/(1-\mu_{\rm c})] (m_{\rm e}c^2)^2/E_i$.
The photon-photon collision is assumed to be isotropic.
Since the typical energy of the curvature photons is around TeV,
only the ADAF photons above near-IR frequency contribute for pair creation.
However, at $\dot{m}<10^{-4}$, the number flux of these photons 
is typically $8$ orders of magnitude smaller than 
the sub-millimeter-IR photons, 
which will effectively collide with $\sim 100$~TeV IC photons.
Thus, for low-luminosity 
active galactic nuclei (AGNs), curvature photons do not effectively
materialize as pairs in the BH gap,
forming a striking contrast to pulsar gaps.
This point will be discussed in \S~\ref{sec:app_IC310} in detail.

\subsection{Gap closure condition}
\label{sec:closure}
Although the curvature photons do not materialize in a BH gap
of a low-luminosity AGN,
we consider both the curvature and IC processes for completeness. 
The number of pairs cascaded from a single electron 
is given by
\begin{equation}
  {\cal M}
  = \Sigma_i \left[ (N_{\rm cv})_i +(N_{\rm IC})_i \right]
             \left[ 1-\exp(-\tau_i) \right] ,
  \label{eq:MP}
\end{equation}
where $i$ denotes the photon energy bin.
For simplicity, we assume that the same $\cal M$ can be applied
for both the outgoing electrons and the ingoing positrons.
For a gap to be sustained stationarily,
multiplicity $\cal M$ should be unity.
That is, a single electron materializes into $\cal M$ pairs
on average within the gap.
The returned $\cal M$ positrons emit copious $\gamma$-rays inwards,
${\cal M}^2$ of which materialize as pairs. 
As a result, a single electron cascades into ${\cal M}^2$ electrons
after one \lq round trip'.
We thus obtain ${\cal M}=1$ as the condition for a gap to be
sustained stationarily.
If this inward-outward symmetry breaks down,
we must impose a more complicated gap closure condition 
that the product of the inward and outward
multiplicities should become unity \citep{hiro13}.

From the condition of ${\cal M}=1$, we can update $s_2$.
We solve equation~(\ref{eq:pois_2})
from $s=s_2$ to the inner boundary $s=s_1$ where $E_\parallel$ vanishes.
The inner and outer boundaries, $s=s_1$ and $s=s_2$,
are determined as a free-boundary problem,
because we impose the gap closure condition, $\cal{M}=1$
and $dE_\parallel/ds$ at $s=s_2$ by specifying $j$
through equations~(\ref{eq:pois_2}) and (\ref{eq:rho_linear}).
The updated $E_\parallel(s)$ is then substituted into 
equations~(\ref{eq:Lcv}) and (\ref{eq:Lic})
to update $\gamma(s)$. 
Subsequently, $(N_{\rm CV})_i$ and $(N_{\rm IC})_i$
are updated by 
(\ref{eq:dFdnu_cv}), (\ref{eq:def_nuc}), (\ref{eq:dFcv}),
(\ref{eq:dFdnu_ic}), and (\ref{eq:dFic}).
Then ${\cal M}=1$ (eq.~[\ref{eq:MP}]) updates $s_2$ again.
We iterate this process until quantities converge.

\section{Application to IC 310}
\label{sec:app_IC310}
We apply the method described above to the radio galaxy IC~310.
This nearby active galaxy ($z=0.0189$) has been detected with
{\it Fermi}/LAT \citep{nero10} in high energy gamma-rays 
and with MAGIC \citep{alek11,alek14b} in VHE.
Also in X-rays, non-thermal point-like emissions have been
detected \citep{schwarz92,rhee94,sato05}.

Using the $M-\sigma$ relation,
one can estimate the mass of the central black hole of IC~310 to be
$(2.4 \pm 0.5) \times 10^8 M_\odot$ 
from the velocity dispersion of 
$(229.6 \pm 5.9) \mbox{ km s}^{-1}$ in the host galaxy
\citep{macElroy95,simien02}.
On the other hand, using the fundamental plane of black hole activity
\citep{merloni03},
one obtains a higher mass
$\sim 4 \times 10^8 M_\odot$
from the 2-10~keV X-ray flux of 
$6.4 \times 10^{-12} \mbox{ erg cm}^{-2} \mbox{ s}^{-1}$
\citep{eisen13}
and the 1.7 and 8.4~GHz radio flux density of $0.1$~Jy
\citep{schulz15}.
Thus, we adopt $M= 0.3 M_9$ as the mass of the central BH of IC~310. 
See also \citet{alek14b} for details.

The split-monopole solution specifies both the poloidal and toroidal
components of the magnetic field \citep{michel73}.
However, in the present paper, we treat the magnetic curvature radius,
$\rho_{\rm c}$, as a free parameter to examine the impact of
instantaneous kinks and bends of local magnetic field 
(\S~\ref{sec:IC310_small_Rc}).
As a fiducial value, we adopt $\rho_{\rm c}=r$, because
$r$ gives an estimate of the typical curvature radius of a toroidally 
wound magnetic field line.

Furthermore, we adopt $j=1$ in equation~(\ref{eq:rho_linear}).
In general, the created current can be determined if we solve
the position-dependent pair production from the set of 
Boltzmann equations of $e^\pm$'s and photons.
In addition, the value of $j$ should be consistent with
the global flow pattern,
which may be constrained by force-free simulations
\citep{spit06}.
Nevertheless, it is out of the scope of the present paper
to pursue the appropriate values of $j$ by such arguments.
Thus, in what follows, 
we simply parameterize $\rho/B$, adopting $j=1$.

\subsection{The case of equilibrium magnetic field}
\label{sec:IC310_weakB}
Let us first evaluate the magnetic field strength with
equation~(\ref{eq:B_eq}).
Substituting $\dot{m}= 10^{-4}$ and $M_9=0.3$
into equation~(\ref{eq:B_eq}),
we find the equilibrium magnetic field strength,
$B=B_{\rm eq}= 7.3 \times 10^2$~G.
Thus, equation~(\ref{eq:L_BZ}) gives
$L_{\rm sd} \approx 5.3 \times 10^{41} \mbox{ ergs s}^{-1}$.
We should notice here that the Blandford-Znajek power, 
$L_{\rm sd}$, gives the upper limit of the gap luminosity,
because the BH gap is only liberating a part of the
rotational energy of the black hole.
Near the black hole, the magnetic field tends to be radial
(or split-monopole-like)
due to the inertia of the plasmas and the causality at the horizon
\citep{hiro92,mckinney12}.
If we sum up the outward emission along such (quasi) radial field lines
over the entire null surface,
the resultant photon flux will become more or less isotropic.
In this case, the photon flux to be detected little
depends on the beaming angles of the emitted photons.

In contrast to $L_{\rm gap} < 5.3 \times 10^{41} \mbox{ ergs s}^{-1}$
mentioned just above,
MAGIC detected the isotropic luminosity
of $\sim 2 \times 10^{44} \mbox{ ergs s}^{-1}$ during the flare
\citep{alek14b}.
Thus, we must conclude that the BH gap can reproduce at most 
$0.3$~\% of the observed flare luminosity of IC~310
if $B \le B_{\rm eq}$.
Since the gap will be quenched by the ADAF-provided pairs 
at $\dot{m}>10^{-4}$,
there is no hope to obtain such a large gap luminosity as
$L_{\rm gap} \approx 2 \times 10^{44} \mbox{ ergs s}^{-1}$ 
(eq.~[\ref{eq:Lgap}]),
if we adopted the equilibrium magnetic field for IC~310.

\subsection{Possibility of stronger magnetic field
            around extreme Kerr hole}
\label{sec:IC310_extreme}
To explore the possibility of a stronger $B$, 
let us consider a magnetized plasma
around an extremely rotating BH, 
$r_{\rm g} > a \ge 0.998 r_{\rm g}$ 
\citep{bard70,thorne74}.
If a black hole is extremely rotating,
causality does not allow an additional accretion of plasmas 
that have positive angular momenta, thereby halting accretion 
at the event horizon. 
As a result of this cancellation of the gravity by the centrifugal force, 
a strong magnetic field can be confined by the accumulated, dense plasmas.
For instance, magnetohydrodynamic accretions have a vanishing 
Alfvenic Mach number, $M_{\rm A} \propto B n^{-1/2}$ \citep{came86a,came86b}, 
and hence a diverging plasma density $n$, 
at the event horizon in the limit of $a \rightarrow r_{\rm g}$
\citep{hiro92}. 
In addition, GR MHD simulation shows that 
the magnetic energy becomes approximately 30 times (or 100 times)
greater for $a=0.998 r_{\rm g}$ near the horizon, 
comparing with $a=0.9 r_{\rm g}$ (or $a=0$) case
under similar initial and boundary conditions
\citep{hirose04}.
This enhancement of the magnetic field is due to a plasma compilation
near the horizon. 
The enhanced magnetic field enters well within the ergosphere, 
having a non-negligible radial component in the funnel.
Thus, in the higher latitudes, 
it is capable of extracting the hole's rotational energy efficiently.

To set up the model as simple as possible,
we therefore assume $B \approx 10^4$~G irrespective of $\dot{m}$
under $r_{\rm g} > a \ge 0.998 r_{\rm g}$ in this paper. 
Although equation~(\ref{eq:L_BZ}) gives 
$L_{\rm sd} \approx 10^{44} \mbox{ ergs s}^{-1}$ in this case,
which greatly exceeds the observed jet power by $50$ times,
we expect that such a strong magnetic field,  $B > 14 B_{\rm eq}$, 
can be sustained near the horizon only in a much shorter period
than the jet propagation time scale.
In another word, the BH gap of IC~310 is activated only intermittently
with a small duty cycle, e.g., $\sim 1/50=0.02$.
Although it is important to confirm the feasibility 
of such a strong $B$ around an extremely rotating BH
by a numerical analysis, 
such an investigation is irrelevant to the main subject of 
the present paper.
If $B$ greatly exceeded $10^4$~G, on the other hand,
the BH’s energy and angular momentum 
would be electromagnetically extracted so efficiently 
that the extracted power would exceed the jet power of 
typical AGNs \citep{ghise14}.
On these grounds, we assume that the BH is extremely rotating,
$a=0.998 r_{\rm g}$, 
and that $B \approx 10^4$~G holds during the flare.
For such an extreme Kerr BH, the horizon radius 
$r_{\rm H}$ becomes about a half of the Schwarzshild radius.



\subsection{Gap structure versus accretion rate}
\label{sec:IC310_structure}
If the BH is extremely rotating at $a \approx 0.998 r_{\rm g}$, 
the null surface  is located at radius
$r=r_0 \approx 2.1 r_{\rm H}$
in $0 \le \theta < \pi/2$ (fig.~\ref{fig:side_null})
if $\Omega_{\rm F} \approx 0.3 \omega_{\rm H}$,
irrespective of the field line geometry.
In what follows, unless explicitly mentioned,
we adopt $\theta=15^\circ$ \citep{alek14b},
and assume a split-monopole magnetic field 
(left panel of fig.~\ref{fig:side_null}).
Note that the argument is subject to change
only mildly if we adopt different field line geometry
such as parabolic one (right panel of fig.~\ref{fig:side_null}),
particularly at higher latitudes, $\theta < 30^\circ$.

\begin{figure}
 \epsscale{1.0}
 \plotone{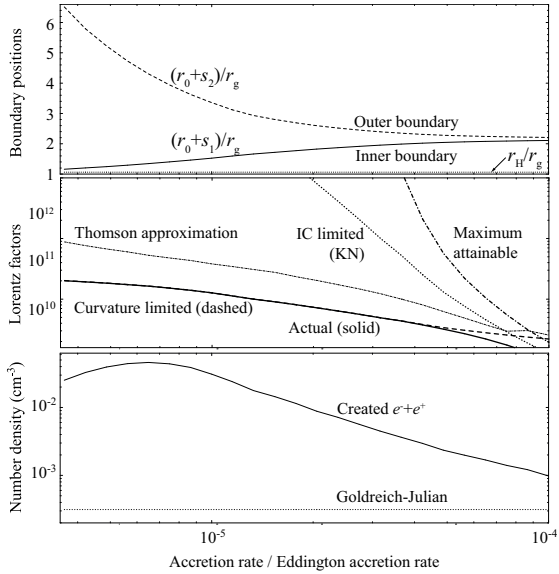}
\caption{
Quantities as a function of the dimensionless accretion rate $\dot{m}$. 
{\it Top panel:}
Boundary positions in Boyer-Lindquist radial coordinate in
$r_{\rm g}$ unit.
The solid and dashed curves show the dimensionless positions of the
inner and outer boundaries, $(r_0+s_1)/r_{\rm g}$ and 
$(r_0+s_2)/r_{\rm g}$. 
The dotted line represents the horizon position,
$r_{\rm H}/r_{\rm g}=1.06$, 
which gives the physical lower limit of $(r_0+s_1)/r_{\rm g}$
(solid curve). 
{\it Middle panel:}
The dash-dotted line shows the maximum attainable Lorentz factor 
computed from the potential drop of the non-vacuum gap. 
The thick dashed and dotted lines do the Lorentz factors limited by the 
curvature radiation and inverse-Compton scatterings, respectively. 
The actual Lorentz factor of the accelerated electrons 
becomes as the solid line. 
For comparison, we also plot the Lorentz factors 
that would be obtained if the IC scatterings
took place in the Thomson regime, as the thin dashed curve.
{\it Bottom panel:}
The solid line shows the density of the created electrons and positrons 
outside the gap, while the dashed one does the Goldreich-Julian value.
\label{fig:results}
}
\end{figure}

The full gap width $w=s_2-s_1$ increases 
with decreasing $\dot{m}$ as depicted 
in the top panel of figure~\ref{fig:results}. 
If the accretion rate decreases to $4.87 \times 10^{-6}$,
$r_0+s_1 \rightarrow r_{\rm H}=1.06 r_{\rm g}$, that is, 
the inner boundary eventually touches down the horizon.
Below this critical accretion rate, no gap solution exists.

In the middle panel, we present
$\gamma_{\rm curv}$ (eq.~[\ref{eq:Lcv}]),
$\gamma_{\rm IC}$ (eq.~[\ref{eq:Lic}]), and
$\gamma$ (eq.~[\ref{eq:Lf}])
with dashed, thick dotted, and solid curves.
Note that the IC drag force is explicitly computed 
by fully taking account of the Klein-Nishina effect.
For reference, we also plot the IC-limited Lorentz factor
$\gamma_{\rm Thomson}$
that would be obtained if the scatterings take place 
in the Thomson regime with thin dotted curve.
It is clear that the IC process takes place in the extreme 
Klein-Nishina regime 
and that the deviation from $\gamma_{\rm Thomson}$ 
becomes prominent at $\dot{m} \ll 10^{-4}$.
Since we compute $\gamma_{\rm Thomson}$ using a single
parameter in the same way as LR11 for reference purpose
(instead of replacing the Klein-Nishina cross section
 with the Thomson cross section in equation~[\ref{eq:Lic}]),
the thick and thin dotted curves do not match around 
$\dot{m} \sim 10^{-4}$.
It also shows that the Lorentz factors are limited by the curvature process
(eq.~[\ref{eq:Lcv}])
in almost the entire range of $\dot{m} < 10^{-4}$.
Since the curvature drag force dominates the IC one, 
the gap luminosity is dominated by the curvature process.
Nevertheless, explicit computation of the IC emission 
with the Klein-Nishina cross section is, indeed, essential.
This is because the closure condition, ${\cal M}=1$,
is satisfied by the IC photons,
or equivalently because the curvature photons do not materialize
as pairs colliding with ADAF soft photons.
In short, the Lorentz factors are limited by $\gamma_{\rm curv}$
in the entire range of $\dot{m}<10^{-4}$ .

In the bottom panel, 
we plot the number density of the pairs created outside the gap 
via photon-photon collisions. 
Since a force-free magnetosphere can be sustained when the created 
pair density exceeds the GJ value, 
the present solutions give natural stationary solutions of the gap. 
The created pair density exceeds the GJ value
more than two order of magnitude when the gap becomes most luminous,
because $\sim 10^4$~TeV IC photons cascade into $~10^2$~TeV pairs.
Since the charge density becomes comparable to the GJ value
in the gap when $w \gg r_{\rm g}$,
it means that the multiplicity for a primary lepton to cascade
into secondaries (and higher generation pairs) is also $\sim 10^2$.
Since we are assuming $B=10^4$~G irrespective of $\dot{m}$,
the GJ density (horizontal dotted line)
is constant with $\dot{m}$ in the present treatment.

\subsection{Gap distribution}
\label{sec:IC310_distribution}
Let us examine the spatial distribution of the gap.
Since the ADAF MeV photon density will be more or less homogeneous,
we apply $\dot{m} = 3.16 \times 10^{-5}$ for all the field lines.
For illustration purpose,
we adopt $\Omega_{\rm F}= 0.3 \omega_{\rm H}$
for all the field lines
as a practical compromise.
It is, however, noteworthy that 
$-0.1\Omega_{\rm H} < \Omega_{\rm F} < 0.35 \Omega_{\rm H}$ is indicated 
in the funnel
by the numerical simulations and the supporting analytical work 
both of which are claimed to represent radio loud AGN
\citep{mckinney12,bes13}.

\begin{figure}
 \epsscale{1.0}
 \plotone{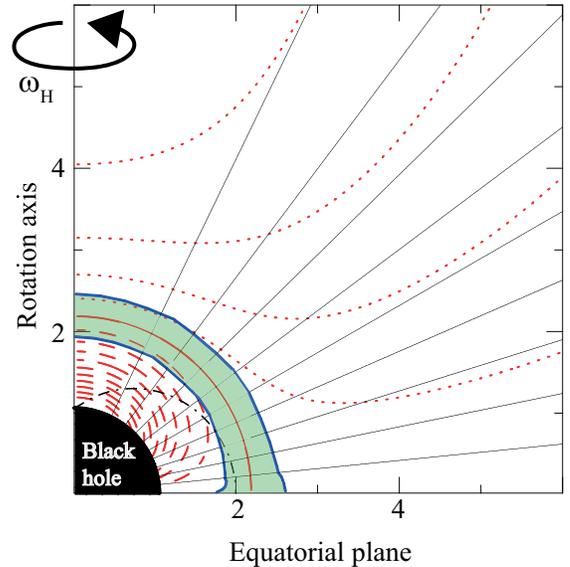}
\caption{
Similar figure as the left panel of fig.~\ref{fig:side_null}
but the gap spatial distribution is depicted as
the green-shaded region when the dimensionless accretion rate is
$\dot{m} = 3.16 \times 10^{-5}$.
The gap solution does not strongly depend on the
magnetic field configuration such as split-monopole
or parabolic ones,
because the null surface position is solely determined
by frame dragging frequency $\omega(r,\theta)$ and
field-line angular frequency $\Omega_{\rm F}(A_\varphi)$.
For example, if $\Omega_{\rm F}$ is constant for all the 
magnetic field lines, the null surface distribute
nearly spherically as depicted.
As $\dot{m}$ decreases further,
the inner boundary (thick blue curve that is closer to the black hole)
shifts further inwards,
and the outer boundary (thick blue curve located away from the hole)
shifts further outwards.
The split-monopole approximation of the magnetic field lines 
may be broken down in the lower latitudes,
because there is an equatorial accretion flow.
\label{fig:sideview}
}
\end{figure}

At $\dot{m} = 3.16 \times 10^{-5}$,
the gap distributes on the meridional plane as depicted by
the green region in figure~\ref{fig:sideview}. 
At $\theta=15^\circ$, $45^\circ$, and $75^\circ$, 
the full gap thickness 
(along the magnetic field line on the poloidal plane)
become $w=s_2-s_1=0.525 r_{\rm g}$, $0.538 r_{\rm g}$, and $0.593 r_{\rm g}$,
respectively (see also the top panel of fig.~\ref{fig:results}).
Since $\Omega_{\rm F}=\Omega_{\rm F}(\Psi)$ is assumed to be constant here,
the null surface distributes nearly spherically.
In this case, 
the gap solution little changes in the higher latitudes, 
because $\rho_{\rm GJ}$ distribution does not have a strong dependence on
$\theta$ in $\theta<30^\circ$. 
Therefore, we assume that the split-monopole magnetic field extends
up to $60 r_{\rm g}$ to compute the pair cascade outside the gap.
This treatment is probably justified in the polar funnel
of accreting BH systems \citep{mckinney07a,mckinney07b}.


\subsection{Electric field distribution}
\label{sec:IC310_Ell}
\begin{figure}
 \epsscale{1.0}
 \plotone{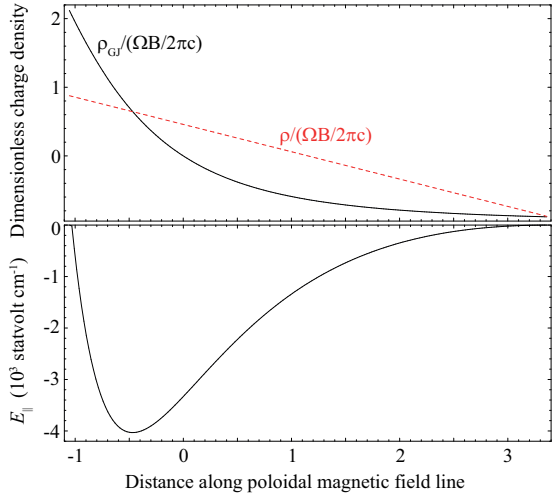}
\caption{
\label{fig:rho_7}
Charge densities and acceleration electric field
in a non-vacuum gap with maximum screening, $j=1.0$,
for the BH magnetosphere of IC~310.
The accretion rate is at the critical value, 
$\dot{m}=4.87 \times 10^{-5}$.
In the abscissa, the inner and outer boundaries are located at
$s_1=-1.07 r_{\rm g}$ and $s_2=3.38 r_{\rm g}$, respectively.
{\it Top panel:} 
Similar figure as the top panel of figure~\ref{fig:rho_2}.
{\it Bottom panel:} 
Similar figure as the bottom panel of figure~\ref{fig:rho_2},
but the ordinate is in $10^3 \mbox{ statvolt cm}^{-1}$ unit.
}
\end{figure}

Computing the real charge distribution with equation~(\ref{eq:rho_linear}),
we obtain $E_\parallel$ for the critical accretion rate,
$\dot{m}=4.87 \times 10^{-6}$,
below which no gap solution exists.
The result is depicted in figure~\ref{fig:rho_7},
where the event horizon is located at $s=-1.09 r_{\rm g}$
on the abscissa.
It follows that $E_\parallel$ maximizes around the middle point
between the horizon and the null surface ($s=0$).
Therefore, a significant fraction of the VHE photons are emitted
within $2 r_{\rm g}<r_0 \approx 2.1 r_{\rm g}$ 
via the curvature process.
The maximum attainable Lorentz factor in the middle panel of 
figure~\ref{fig:results}
is computed from the solved $E_\parallel(s)$ at each $\dot{m}$,
as depicted in the lower panel of figure~\ref{fig:rho_7}.

\subsection{Emission spectra}
\label{sec:IC310_spectra}
\begin{figure}
 \epsscale{1.15}
 \plotone{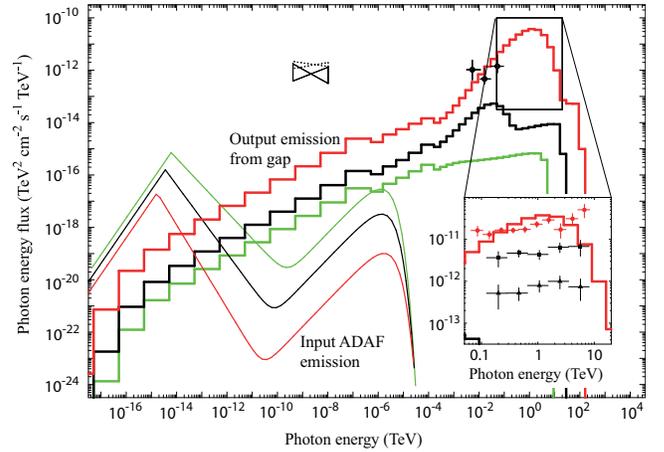}
\caption{
Spectral energy distribution of the photons emitted from the gap 
exerted in the vicinity the supermassive BH of IC 310. 
The thin green, black and red curves show the input ADAF spectra 
\citep{mahad97}
for $\dot{m} =1.0 \times 10^{-4}$, 
$3.16 \times 10^{-5}$, and 
$4.87 \times 10^{-6}$, respectively.
The thick three lines show the output emission of the gap 
at each accretion rate. 
The thick red line gives the maximum photon flux,
because the gap inner boundary almost touches down the horizon
at $\dot{m}=4.87 \times 10^{-6}$.
Below this critical accretion rate, gap solution
ceases to exist.
The magnetic field strength is assumed to be
$B=10^4 (r_0/r)^2$~G at Boyer-Lindquist radius $r$,
where $r_0$ denotes the radius at the null surface.
At greater accretion rate $\dot{m} \sim 1.0 \times 10^{-4}$,
the inverse-Compton process dominates the curvature one;
thus, the thick green line exhibits a power-law spectrum
between 100~MeV and 2~TeV.
The inset shows a close-up in the VHE regime; 
the black triangles and squares 
show the observed fluxes in the low and high states 
\citep{alek11} respectively, 
while the red circles do those in the flaring state \citep{alek14b}. 
The observed VHE fluxes are corrected for the absorption 
by the extragalactic background light.
The solid bowtie represents the X-ray fluxes
obtained by re-analyzing 
the observational data taken with
{\it XMM-Newton} in 2003, 
{\it Chandra} in 2004, and 
{\it Swift} in 2007 \citep{alek14a}.
The dotted bowtie represents the re-analyzed data 
of {\it Chandra} in 2005 \citep{alek14a}.
The open circles denote the {\it Fermi}/LAT spectrum.
\label{fig:SED1}
}
\end{figure}

In figure~\ref{fig:SED1}, 
we depict the spectra of the photons emitted by the 
out-going electrons as thick lines, 
for three discrete accretion rate, $\dot{m}$.
The out-going particles consist of 
the primary electrons accelerated in the gap
and the cascaded pairs outside the gap.
These particles up-scatter the background soft photons,
which will be partly absorbed by the same soft photon field.
The soft photons are provided by the ADAF, 
whose corresponding spectra \citep{mahad97}
are indicated by the thin curves. 
In the $\dot{m} \rightarrow 0$ limit,
$dF_{\rm s}/dE_{\rm s} \rightarrow 0$ means 
$w =s_2-s_1 \rightarrow \infty$
in equation~(\ref{eq:tau})
in order that ${\cal M}=1$ may be held in equation~(\ref{eq:MP}).
Thus, $r_0+ s_1$ should approach the lower limit, $r_{\rm H}$, 
if $\dot{m}$ reduces enough. 
In the present case, $r_0+s_1$ reaches $r_{\rm H}$, 
when $\dot{m}=4.87 \times 10^{-6}$
(upper panel of fig.~\ref{fig:results}).
This most luminous case is depicted as the red thick line
in figure~\ref{fig:SED1}. 
We find that the flux of the flare 
(red filled circles in the inset) can be qualitatively reproduced.
It looks contradictory that the gap becomes more luminous 
with decreasing accretion rate.
However, it is a natural consequence of rotation-powered
particle accelerator such as in pulsars.
In the present case, 
it is the BH's rotational energy that energizes the gap.
It forms a striking contrast to accretion-powered systems.

Figure~\ref{fig:SED1} also shows
that the VHE spectrum extends into higher energies
with increasing VHE flux.
This is because the soft photon density decreases
with decreasing $\dot{m}$, 
thereby reducing the absorption of the primary IC photons above 10 TeV. 
Owing to this photon-photon collisions,
there appears a sharp cutoff around 50-100~TeV.
In the most luminous case (red solid line),
the cutoff appears around 100~TeV.

{\it Fermi}/LAT observed IC~310 between August 5, 2008 and July 31, 2011,
and detected eight photons between 12~GeV and 148~GeV \citep{alek14a};
the fluxes are plotted by the open circles.
Considering the small duty cycle of the IC~310's BH gap, 
the {\it Fermi}/LAT flux, which was averaged over the three years,
appears to be higher than our prediction.
In addition, the X-ray fluxes, which are represented by the solid and
dotted bowties, obviously exceed our prediction.
The strong X-ray flux may be due to the emission from 
a corona or the jet.

\begin{figure}
 \epsscale{1.0}
 \plotone{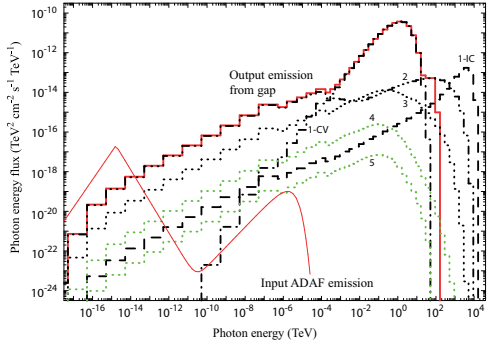}
\caption{
Similar figure as figure~\ref{fig:SED1}, but only for the case of 
the critical accretion rate, $\dot{m} = 4.87 \times 10^{-6}$,
is depicted. 
The thin and thick red solid lines are identical with those in fig. 2. 
The thick dashed line (labeled by 1-IC) represents 
the flux of the inverse-Compton (IC) 
photons emitted by the primary electrons,
while the dash-dotted line (labeled by 1-CV) represents that of 
the curvature photons emitted by the primary electrons.
The dash-dot-dot-dotted line (labeled by 2) shows the sum of
the IC and the synchrotron photons emitted by the secondary pairs 
created outside the gap.
The black dotted line (labeled by 3) 
represents the tertiary photon spectrum,
while the green dotted ones (labeled by 4 and 5)
the quaternary and quinary components.
%
\label{fig:SED2}
}
\end{figure}

In figure~\ref{fig:SED2},
we plot the total, primary IC, primary curvature spectra
as the thick red solid, black dashed, and black dash-dotted lines
for the case of critical accretion rate, $\dot{m}=4.87 \times 10^{-6}$;
the red lines are common in figures~\ref{fig:SED1} and \ref{fig:SED2}.
Since the mean-free path of IC scatterings is much longer than $w$,
the highest energy photons ($>100$~TeV) are mostly emitted 
outside the gap. 
On the contrary, 
since the electrons cool down at much shorter interval than $w$ 
via the curvature process, 
the curvature photons are mostly emitted inside the gap. 
These curvature photons dominates the 0.1-10~TeV flux,
becoming more and more important with decreasing $\dot{m}$,
and hence with increasing output flux.
This is because the curvature emission power is proportional to
$\gamma^4$, while the IC one to $\gamma^1 \sim \gamma^2$
(depending on whether the scatterings take place 
 in the extreme Klein-Nishina or the Thomson limit).
Since the absorption works mainly above 10~TeV,
the curvature photons are little absorbed by the ADAF soft photon field.

The primary IC photons are, on the contrary, heavily absorbed
to be reprocessed as the secondary synchrotron and IC
components, which are depicted as the dash-dot-dot-dotted line.
These secondary photons are absorbed again above 10~TeV
to be reprocessed as the tertiary component (black dotted line).
We compute such a pair-creation cascade until the 30th generation,
depicting the quaternary and quinary components
as the green dotted lines (as labeled with 4 and 5).

\begin{figure}
 \epsscale{1.15}
 \plotone{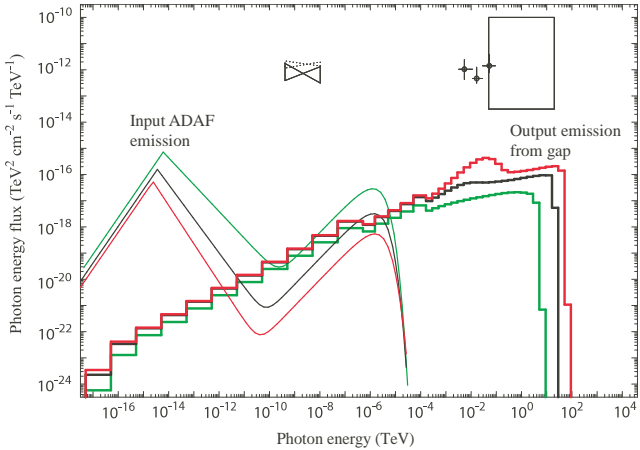}
\caption{
Similar figure as fig.~\ref{fig:SED1}
but the equilibrium magnetic field strength is used
(\S~\ref{sec:IC310_weakB}),
instead of $B=10^4 (r_0/r)^2$~G.
The green and black thin curves represent the input ADAF
spectra for 
$\dot{m}=1.00 \times 10^{-4}$ and $\dot{m}=3.16 \times 10^{-5}$,
respectively,
whereas the red thin curve does that for 
$\dot{m}=5.62 \times 10^{-6}$.
\label{fig:SED1_eq}
}
\end{figure}

Let us briefly examine the case in which $B$ is evaluated 
by its equilibrium value, equation~(\ref{eq:B_eq}).
In figure~\ref{fig:SED1_eq}, we present the photons spectra.
The input parameters are the same as figure~\ref{fig:SED1}.
However, the critical accretion rate becomes
$\dot{m}=5.62 \times 10^{-6}$ in this case.
As analytically expected in \S~\ref{sec:IC310_weakB},
the weaker magnetic field, $B=43.3$~G at this critical accretion rate, 
cannot reproduce
the observed VHE fluxes at all.

It may be desirable to sum up the main points
that have been made in this subsection \ref{sec:IC310_spectra}.\\ 
(1a) The primary electrons (or positrons) are accelerated
outward (or inward) by the magnetic-field-aligned electric field.\\
(1b) The primary particles emit primary photons efficiently 
via the IC and the curvature processes.\\
(1c) If these primary photons materialize as pairs in the gap,
the created pairs are separated and accelerated to become the primary
particles as listed in item (1a).\\
(1d) The observed VHE flare photons are mostly emitted 
via the curvature process.\\
(1e) Although the curvature process dominates the IC one,
it is the IC photons that materialize within the gap,
colliding with the ADAF submillimeter photons.
Thus, both curvature and IC processes are essential,
forming a striking contrast to pulsar gap models,
in which only the curvature process governs the electrodynamics
(because of the higher soft photon energies).\\
(2a) If these primary photons materialize outside the gap,
the created pairs migrate away from the gap as the secondary pairs.\\
(2b) These secondary pairs emit secondary photons 
via the IC and the synchrotron processes.\\
(3a) If these secondary photons materialize within the magnetosphere,
the created pairs migrate away from the gap as the tertiary pairs.\\
(3b) These tertiary pairs emit tertiary photons 
via IC and synchrotron processes, and so on.\\
(4) We recur such calculations 
until the pairs or the photons propagate to $60 r_{\rm g}$,
or until the pairs cascade into the 30th generation.

%
%

\subsection{Created pairs outside gap}
\label{sec:IC310_pairs}
\begin{figure}
 \plotone{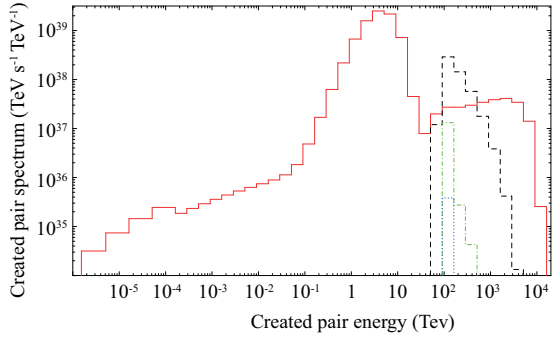}
\caption{
Differential pair creation rate within the volume $r<60r_{\rm g}$.
The red solid, black dashed, green dash-dotted, and 
blue dotted, 
lines represent the creation rate of the 
secondary, tertiary, quaternary, and quinary
generation pairs.
%
\label{fig:pairs}
}
\end{figure}

Let us look in more details at the pairs cascaded outside the gap.
To examine the most efficient case of pair cascade,
we consider the solution at the critical accretion rate,
$\dot{m}=4.87 \times 10^{-6}$ for $B=10^4 (r_0/r)^2$~G
(i.e., the red curves in figs.~\ref{fig:SED1} \& \ref{fig:SED2}).
Figure~\ref{fig:pairs} shows the pairs' energy spectra,
integrated over the entire volume of the BH magnetosphere
within $r_0+s_2<r<60r_{\rm g}$, assuming a spherical symmetry.
The primary IC photons collide with the radio-submillimeter
ADAF photons to materialize as the secondary pairs
whose energy spectrum is represented by the red solid line.
Such secondary pairs up-scatter the ADAF soft photons
that are capable of materializing into the tertiary pairs
(black dashed).
Such tertiary pairs emit tertiary $\gamma$-rays,
some of which materialize as the quaternary pairs
(green dash-dotted).
At such a low accretion rate as 
$\dot{m}=4.87 \times 10^{-6}$,
pair creation cascade finishes at the quinary generation
(blue dotted).
The spectrum sharpens with increasing generation.
The reasons are twofolds:\\
(i) The energy of the cascading pairs decreases with 
increasing generation.\\
(ii) The higher generation pairs are created at greater $r$
where the ADAF photon intensity reduces by $r^{-2}$ law.
Thus, only relatively higher energy pairs can cascade
into higher generations,
because lower energy pairs can emit only the lower energy IC photons
that cannot materialize at greater $r$.\\
In this way, 
the energy range is restricted from both higher and lower sides,
showing a peaking profile around 100~TeV
above the quaternary generation.

Typically speaking, primary IC photons around $10^4$~TeV
cascade into $10^2$~TeV pairs, which means that 
a single primary electron eventually cascade 
into $\sim 10^2$ pairs in the BH magnetosphere.
See also LR11 for similar conclusions.

Integrating over the pair energies,
we obtain $3.03 \times 10^{41} \mbox{ ergs s}^{-1}$
as the kinetic luminosity of the created pairs.
Since the primary ICS luminosity is
$6.97 \times 10^{41} \mbox{ ergs s}^{-1}$,
$43$\% of the up-scattered photon energy 
is converted into cascaded pairs outside the gap.
However, it is only $0.12$\% of the primary curvature luminosity.

\subsection{Dependence on curvature radius}
\label{sec:IC310_small_Rc}
\begin{figure}
 \epsscale{1.15}
 \plotone{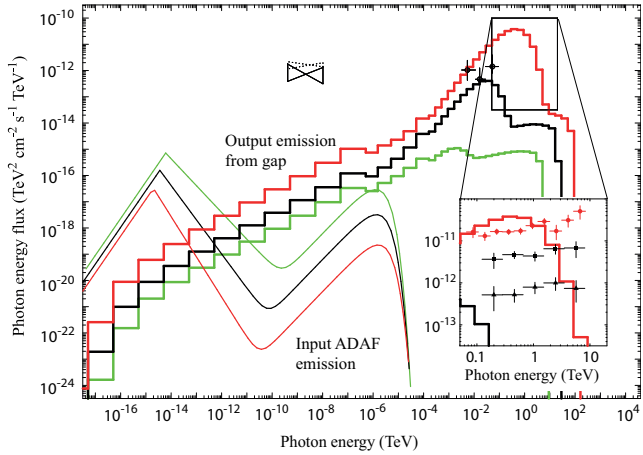}
\caption{
Similar figure as figure~\ref{fig:SED1} 
but a smaller curvature radius, $\rho_{\rm c}=0.1r$,
is assumed at radius $r$.
The green, black and red curves corresponds to the case of
$\dot{m} =1.00 \times 10^{-4}$, 
$3.16 \times 10^{-5}$, and 
$7.50 \times 10^{-6}$
(critical accretion rate), respectively.
\label{fig:SED3}
}
\end{figure}

\begin{figure}
 \epsscale{1.15}
 \plotone{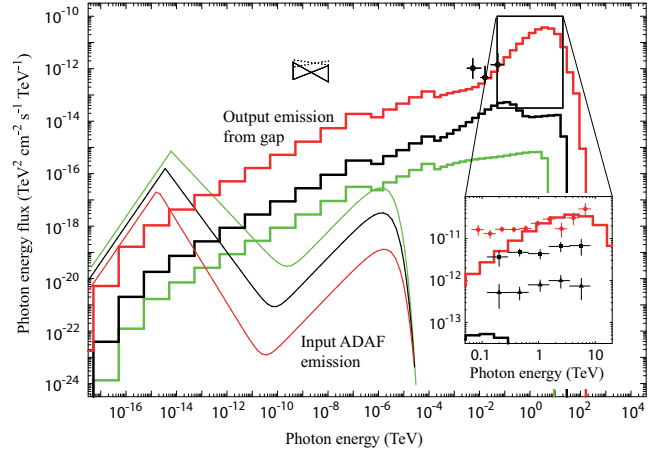}
\caption{
Similar figure as figure~\ref{fig:SED1} 
but a greater curvature radius, $\rho_{\rm c}=10r$, is assumed.
The green, black and red curves corresponds to the case of
$\dot{m} =1.00 \times 10^{-4}$, 
$3.16 \times 10^{-5}$, and 
$5.62 \times 10^{-6}$, respectively.
\label{fig:SED4}
}
\end{figure}

Let us investigate how the solution depends on the
curvature radius, $\rho_{\rm c}$.
In figures~\ref{fig:SED3} and \ref{fig:SED4},
we plot the spectrum obtained for
$\rho_{\rm c}=0.1r$ and $10r$ at radius $r$,
respectively.
The green and black curves represent the cases of 
$\dot{m} =1.00 \times 10^{-4}$, 
$3.16 \times 10^{-5}$, in the same manner as in figure~\ref{fig:SED1}.
The red curve corresponds to the critical accretion rate;
$\dot{m} =7.50 \times 10^{-6}$ and  
$\dot{m} =5.62 \times 10^{-6}$
for figures~\ref{fig:SED3} and \ref{fig:SED4}, respectively.
It follows that the curvature spectrum,
which peaks between 0.1 and 1~TeV, 
becomes softer with decreasing $\rho_{\rm c}$.
This is because equations~(\ref{eq:Lcv}) and (\ref{eq:def_nuc}) give
$\nu_{\rm c} \propto \rho_{\rm c}{}^{1/2}$,
which shows that the curvature-emitted photon energy
decreases an order of magnitude
if $\rho_{\rm c}$ decreases from $10r$ to $0.1r$.
The inset of figure~\ref{fig:SED3} shows that the 
flare spectrum is difficult to be reproduced above 1~TeV,
if the curvature radius becomes as small as $\sim 0.1 r$.
In this case, electrons' Lorentz factors saturate
at one order of magnitude smaller values than the dashed curve
in the middle panel of figure~\ref{fig:results},
because $E_\parallel$ in equation~(\ref{eq:Lcv}) 
depends on $\rho_{\rm c}$ only weakly in the vicinity of a rotating BH,
unlike the situation around a rotating neutron star.
Note that $\partial\rho_{\rm GJ}/\partial s$ is due to
the frame-dragging effect in BH magnetospheres,
whereas it is due to the global field line curvature in
neutron-star magnetospheres.
Thus, $E_\parallel$ little depends on $\rho_{\rm c}$ in BH magnetospheres.


In a time-averaged sense, in the polar funnel, 
the magnetic field lines threading the horizon are 
well-ordered and approximated by a split-monopole configuration
\citep{mckinney07a,mckinney12}.
Although the split-monopole field is tightly wound
azimuthally due to frame dragging near the horizon,
it is predominantly radial with loosely wound helics
moderately away from the horizon (e.g., at $r>2 r_{\rm g}$)
\citep{hirose04}.
Thus, we consider $\rho_{\rm c} \sim 10r$ may be possible
in a time-averaged sense.

Despite the constancy of time-averaged magnetized accretion,
the instantaneous accretion rate varies considerably with time.
In addition, the instantaneous magnetic field lines
are kinked and bent in the evacuated funnel,
which becomes a cauldron of strong waves
\citep{krolik05}.
Thus, it may be reasonable to consider a superposition of
the spectra obtained under different $\rho_{\rm c}$'s.
For instance, due to a fluctuating magnetic field structure,
photons may be emitted into various directions from 
each position in a time-dependent manner.
In such a case, we will detect the combined spectra from
different field lines with various $\rho_{\rm c}$'s.

\begin{figure}
 \epsscale{1.15}
 \plotone{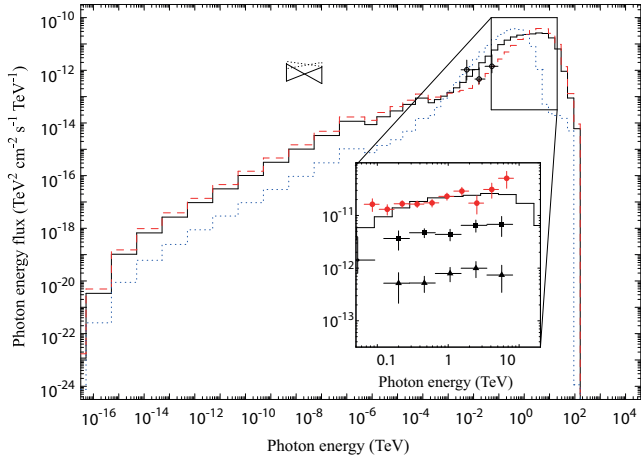}
\caption{
Similar figure as figure~\ref{fig:SED1} 
but the emission spectra with different curvature radius are
superposed.
The red dashed line shows the spectral energy distribution
for $\rho_{\rm c}=10r$, whereas 
the blue dotted one does one 
for $\rho_{\rm c}=0.1r$.
The black solid line shows the superposition of these two curves
with weight $0.75$ and $0.25$, respectively.
In the inset, only the solid line is depicted for clarity.
Note that different $\dot{m}$ is adopted for the dashed and dotted lines.
See text for details.
\label{fig:SED5}
}
\end{figure}

As an example, in figure~\ref{fig:SED5},
we present the result for the superposition of the 
$\rho_{\rm c}=0.1r$ (blue dotted) and 
$\rho_{\rm c}=10r$ (red dashed) cases
with weight $0.25$ and $0.75$, respectively.
Since the ADAF photon field may be inhomogeneous
\citep{li08},
we adopt here different $\dot{m}$ for the 
$\rho_{\rm c}=0.1r$ and $10r$ cases; namely, 
$\dot{m} =7.50 \times 10^{-6}$ for the blue dotted line, and 
$\dot{m} =5.62 \times 10^{-6}$ for the red dashed line.
That is, the blue dotted line coincides with the red solid line
in figure~\ref{fig:SED3}, whereas
the red dashed line does the red solid line
in figure~\ref{fig:SED4}.
It follows that a power-law-like spectrum can be formed
if curvature spectra with varying $\rho_{\rm c}$
are superposed with appropriate weights.
So that the observed power-law-like VHE spectrum may be reproduced,
greater weights are favorable at greater $\rho_{\rm c}$'s.
In another word,
the observed flare spectrum may imply 
that the instantaneous magnetic field lines
are mostly straight but kinked moderately in the funnel.
The actual distribution of the weight as a function of $\rho_{\rm c}$
should be examined separately by numerical analysis.
However, it is out of the scope of the present paper.

\subsection{Dependence on magnetic-field angular frequency}
\label{sec:IC310_omgF}
\begin{figure}
 \epsscale{1.15}
 \plotone{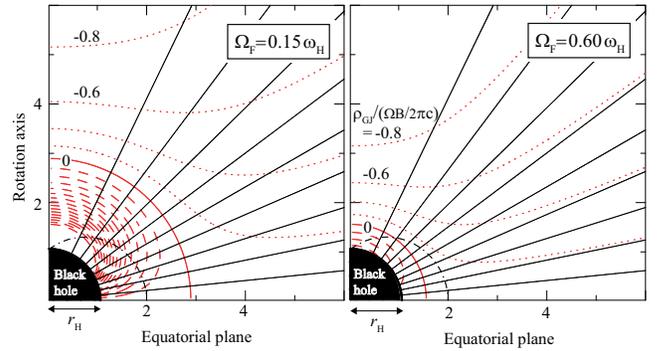}
\caption{
Similar figure as the left panel of figure~\ref{fig:side_null}:
both panels show the dimensionless Goldreich-Julian charge density 
(red dashed, solid, and dotted contours) 
for a split-monopole magnetic field (black solid lines).
The left panel is for a slower field-line rotation, 
$\Omega_{\rm F}=0.15 \omega_{\rm H}$,
while the right panel for a faster rotation,
$\Omega_{\rm F}=0.60 \omega_{\rm H}$.
\label{fig:side_null2}
}
\end{figure}

Let us finally examine the dependence on $\Omega_{\rm F}$.
First, we consider the distribution of the dimensionless 
GJ charge density, $\rho_{\rm GJ}/(\Omega_{\rm F} B/2\pi c)$,
on the poloidal plane,
assuming a split-monopole magnetic field line configuration.
In the left panel of figure~\ref{fig:side_null2},
we adopt a slower field-line rotation,
$\Omega_{\rm F}=0.15 \omega_{\rm H}$.
It follows that the null surface (red solid curve)
shifts to $r \approx 2.9 r_{\rm g}$ from $\approx 2.1 r_{\rm g}$
as $\Omega_{\rm F}$ reduces to $0.15 \omega_{\rm H}$
from $0.30 \omega_{\rm H}$ (cf. left panel of fig.~\ref{fig:side_null}).
This is because $\omega$ reduces by $r^{-3}$, which indicates that
$\omega=\Omega_{\rm F}$ is realized at greater $r$
for a smaller $\Omega_{\rm F}$.
We plot the (red) dashed contours 
only below $\rho_{\rm GJ}/(\Omega_{\rm F} B/2\pi c)=2.0$,
because the gap inner boundary is located in the
region where $\rho_{\rm GJ}/(\Omega_{\rm F} B/2\pi c)<2.0$ holds,
unless a super-GJ electric current is injected across the outer boundary.
On the other hand, if the magnetic field lines rotate much faster
as $\Omega_{\rm F}=0.60 \omega_{\rm H}$,
the null surface is located at $r \approx 1.5 r_{\rm g}$,
as the right panel shows.
In this case, the inner boundary more easily touches down the horizon,
because the GJ charge density at the horizon, 
$\vert \rho_{\rm GJ} \vert_{\rm H}$ 
is proportional to $\omega_{\rm H}-\Omega_{\rm F}$,
thereby being limited at a smaller value 
if $\Omega_{\rm F}$ approaches $\omega_{\rm  H}$.

\begin{figure}
 \epsscale{1.15}
 \plotone{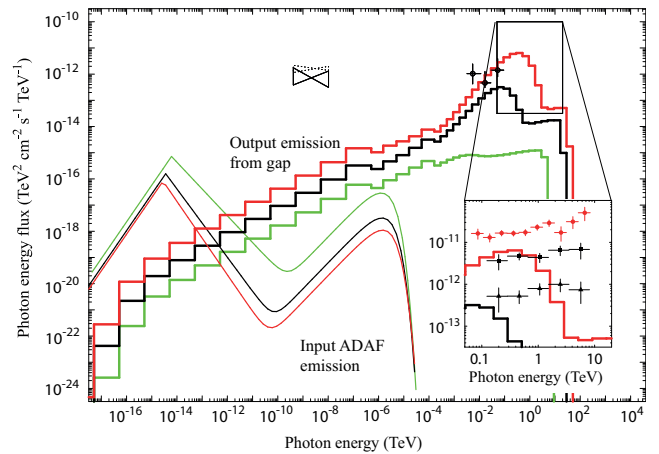}
\caption{
Similar figure as figure~\ref{fig:SED1} 
but a greater rotational angular frequency, 
$\Omega_{\rm F}=0.6 \omega_{\rm H}$, is assumed for the magnetic field.
The green, black and red curves corresponds to the case of
$\dot{m} =1.00 \times 10^{-4}$, 
$3.16 \times 10^{-4}$, and 
$1.78 \times 10^{-5}$
(critical accretion rate), respectively.
\label{fig:SED6}
}
\end{figure}

Second, we consider the $\gamma$-ray spectrum.
At $\Omega_{\rm F} < 0.30 \omega_{\rm H}$,
we can obtain a similar VHE spectrum as the $0.30 \omega_{\rm H}$ case
when the inner boundary does not touches down the horizon,
provided that $B$ is unchanged at $10^4$~G in the gap.
Alternatively, we obtain a similar VHE spectrum
for $B<10^4$~G when the inner boundary touches down the horizon.
On the other hand, at $\Omega_{\rm F} > 0.30 \omega_{\rm H}$,
we cannot obtain the observed flare VHE flux 
even when the inner boundary touches down the horizon,
provided that $B=10^4$~G.
We could adopt $B>10^4$~G and obtain a similar VHE spectrum;
however, a greater $B$ would result in a further greater
jet-power efficiency (see \S~\ref{sec:limitations}),
which is unfafoured.
In figure~\ref{fig:SED6}, we present the spectra
obtained for $\Omega_{\rm F}=0.60 \omega_{\rm H}$
and $B=10^4$~G.
It shows that the VHE flux reduces significantly
and that the curvature spectrum softens 
if $\Omega_{\rm F} > 0.3 \omega_{\rm H}$ 
(cf. fig.~\ref{fig:SED1}).
Thus, we can conclude that a smaller value of $\Omega_{\rm F}$
in the polar funnel, which is suggested from numerical and 
analytical works \citep{mckinney12,bes13},
is favourable to obtain a greater VHE flux from a BH gap.

\section{Discussion}
\label{sec:disc}
In summary, a gap (i.e., a low plasma density region) arises 
around the null-charge surface that is formed by the frame-dragging 
effect around a rotating black hole (BH).
The gap width along the magnetic field lines, 
and hence its luminosity increases with
decreasing plasma accretion from the surroundings.
In the case of IC 310, the observed jet luminosity
and the inferred BH mass give $\dot{m} \approx 2 \times 10^{-4}$
as the time-averaged, dimensionless mass accretion rate.
However, at this accretion rate, the radiatively inefficient
accretion flow supplies the electron-positron pairs whose 
density exceeds the Goldreich-Julian value by an order of magnitude.
Thus, in a large fraction of time, the BH gap is expected to be
quenched around the supermassive BH of IC310.
Nevertheless, the strong dependence of the created pair density
by ADAF photon field (eq.~[\ref{eq:n_ratio}])
implies the activation of a BH gap with a small duty cycle
by virtue of a time-dependent plasma accretion.
In the case of IC~310, it is expected that a BH gap is 
activated by a charge deficit
if the accretion rate is halved from the time-averaged value.
If the  magnetic field is in equilibrium with the plasma's
gravitational biding energy, we obtain $B<10^3$~G for IC~310.
However, the predicted gamma-ray flux turns out to be 
more than two orders of magnitude smaller than the flare flux.
Thus, to contrive a stronger magnetic field strength,
we assume an extremely rotating BH, $a=0.998 r_{\rm g}$.
Provided that $B \approx 10^4$~G is maintained intermittently
near the horizon,
we find that the flare flux can be reproduced 
when the gap inner boundary nearly touches down the horizon.
The TeV spectrum is predicted to extend into higher energies
with increasing TeV flux.


\subsection{Limitations of the present model when applied to IC~310}
\label{sec:limitations}
\subsubsection{Energy extraction efficiency}
\label{sec:efficiency}
When the accretion rate takes its time-averaged value,
$\dot{m} \approx 2 \times 10^{-4}$, 
the radiation inefficient accretion flow (RIAF) 
supplies copious photons via the free-free process.
These MeV photons collide each other so efficiently
that the created pair density exceeds the Goldreich-Julian (GJ) value.
In this case, the hole's rotational energy is extracted via
the Blandford-Znajek process 
(i.e., the electromagnetic energy extraction process in a force-free
magnetosphere) and a steady jet is emitted with luminosity 
$L_{\rm jet} \approx 2 \times 10^{42} \mbox{ ergs s}^{-1}$,
which is consistent with the radio observations (\S~\ref{sec:RIAF}).
During this jet phase, we assume that the magnetic field 
takes the equi-partition value~(eq.~[\ref{eq:B_eq}]) 
in a time-averaged sense.

On the contrary, 
when the accretion rate becomes $\dot{m} < 10^{-4}$,
the magnetosphere becomes charge-starved and the BH gap is switched on.
To explain the enormous flare luminosity, 
$L \approx 2 \times 10^{44} \mbox{ ergs s}^{-1}$,
we assumed in the present paper that the
magnetic field increases about ten times (to $B \approx 10^4$~G) 
intermittently near the null surface.
Note that $B$ does not always have to become $\approx 10^4$~G
when $\dot{m}$ reduces.
Instead, a local and intermittent increase of $B$ to 
$\approx 10^4$~G (coincidentally when $\dot{m} \ll 10^{-4}$) 
is sufficient for the BH gap to exhibit a flare activity.
The extracted power attains $\sim 100$ times greater value
than the accretion power (eq.~[\ref{eq:B_eq}]),
which means an efficiency of $\sim 10^4$\%.
However, such a high extraction efficiency has not been 
demonstrated by any numerical simulations so far.
For example, 
\citet{mckinney12} considered a rapidly rotating BH,
$a \approx 0.9375 r_{\rm g}$, 
and showed that the jet efficiency could attain at most
$\sim 900$\%. 
Thus, by numerical techniques,
it is necessary to demonstrate 
if such an intermittent compilation of a magnetized plasma
($B\sim 10^4$~G),
and the resultant very high efficiency ($\sim 10^4$\%), 
is indeed possible 
around an extremely rotating BH ($a=0.998 r_{\rm g}$).

Unless one considers a highly anisotropic emission
(e.g., by relativistic beaming), 
any emission models will encounter the same problem
of a huge efficiency ($\sim 10^4$\%).
It may indicate that the models refuted in
\citet{alek14b}, 
such as the jet-in-jet model 
\citep{giannios10} 
or the jet-cloud interaction model 
\citep{bednarek97,barkov10,barkov12},
may be worth reconsidering.

\subsubsection{Photon density constrained by X-ray observations}
\label{sec:X-ray}
In this paper, we have examined the ADAF theory and 
concluded that the BH magnetosphere 
of IC~310 becomes charge starved if $\dot{m}<10^{-4}$.
To check this logic, in this subsection,
we compare the MeV photon density predicted by ADAF
(eq.~[\ref{eq:ns}]),
\begin{equation}
  n_{\rm s}
  = 5.64 \times 10^4 
    \left( \frac{M_9}{0.3} \right)^{-1}
    \left( \frac{\dot{m}}{2 \times 10^{-4}} \right)^2
    \left( \frac{r}{6 r_{\rm g}} \right)^{-2}
    \mbox{ cm}^{-3} ,
  \label{eq:n_gamma}
\end{equation}
with observations.

Unfortunately, there is no direct observations around MeV
for IC~310 along with other low accretion rate BH systems.
Thus, to estimate the MeV photon density,
here we utilize the X-ray observations
and extrapolate the spectrum into MeV energies.

The X-ray spectrum of IC~310 has been obtained between 
2~keV and 10~keV with 
{\it XMM-Newton} (MJD 52697, i.e., February 27, 2003),
{\it Chandra} (MJD 53363, 53456), and
{\it Swift-XRT} (MJD 54152)
\citep{alek14a,sato05}. 
The highest X-ray flux 
$2.5 \times 10^{-3} \mbox{ keV s}^{-1} \mbox{ cm}^{-2}$
(with photon index $\Gamma=2.01$)
was obtained by 
{\it Chandra} (MJD 55456), while the lowest one 
$8.28 \times 10^{-4} \mbox{ keV s}^{-1} \mbox{ cm}^{-2}$
(with photon index $\Gamma=2.55$)
by {\it XMM-Newton}.
We will consider the both cases in this subsection
to estimate the lower and upper bound of the photon number density
around MeV.

First, to extrapolate the observed power-law spectrum into MeV,
we assume an exponential cutoff above energy $E_{\rm br}$, 
in the differential photon number flux,
\begin{equation}
  \frac{df}{dE}
  = N E^{-\Gamma} \exp{\left(-\frac{E}{E_{\rm br}}\right)}
    \mbox{ s}^{-1} 
    \mbox{cm}^{-2} 
    \mbox{keV}^{-1},
  \label{eq:dFdE}
\end{equation}
where the photon energy $E$ is measured in keV.
Multiplying $E$ and integrating equation~(\ref{eq:dFdE}) 
from 2 to 10~keV,
we obtain the energy flux in $2-10$~keV, 
\begin{equation}
  F_{\rm X}= N \frac{10^{2-\Gamma}-2^{2-\Gamma}}
                   {2-\Gamma}
    \mbox{keV s}^{-1} 
    \mbox{cm}^{-2}. 
  \label{eq:dFdE_2_10}
\end{equation}
Setting 
 $F_{\rm X}=2.5 \times 10^{-3} \mbox{ s}^{-1}
                             \mbox{cm}^{-2}
                             \mbox{keV}^{-1} $ 
and
$8.28 \times 10^{-4} \mbox{ s}^{-1}
                    \mbox{cm}^{-2}
                    \mbox{keV}^{-1} $, 
we obtain 
\begin{equation}
  N=1.55 \times 10^{-3} 
  \label{eq:N_1},
\end{equation}
\begin{equation}
  N=1.13 \times 10^{-3} 
  \label{eq:N_2},
\end{equation}
for the {\it Chandra} (MJD 55456) and {\it XMM} observations, respectively.
We estimate the photon number density at MeV by
\begin{equation}
  n_{\rm s,obs} 
  \equiv \left( \frac{d}{r} \right)^2
         \frac{1}{c}
         \left( E \frac{df}{dE} \right)_{\rm MeV},
  \label{eq:n_gamma_obs}
\end{equation}
where $d=81.4$~Mpc denotes the distance to IC~310, and
\begin{equation}
  \left( E \frac{df}{dE} \right)_{\rm MeV}
  = 10^{-3(\Gamma-1)} N \exp{\left(-\frac{E}{E_{\rm br}}\right)}
    \mbox{s}^{-1} \mbox{ cm}^{-2}
  \label{eq:dFdE_MeV}
\end{equation}
denotes the photon number flux at MeV.
Thus, the photon number density at $r=6r_{\rm g}$ can be estimated to be
\begin{equation}
  n_{\rm s,obs} 
  = 2.97 \times 10^{16-3\Gamma}
      N e^{-E/E_{\rm br}} (r/6r_{\rm g})^{-2}
      \mbox{ cm}^{-3} .
  \label{eq:n_gamma_obs_0}
\end{equation}

For the low X-ray flux case of {\it XMM} observation,
we obtain
\begin{equation}
  n_{\rm s,obs} 
  = 8.8 \times 10^5 e^{-E/E_{\rm br}} (r/6r_{\rm g})^{-2}
    \mbox{ cm}^{-3}.
  \label{eq:n_gamma_obs_1}
\end{equation}
Therefore, we find that the e-folding energy $E_{\rm br}$
should be less than 370~keV 
so that $n_{\rm s,obs}$ may become less than $n_{\rm s}$ 
at the time-averaged accretion rate, 
$\dot{m}= 2 \times 10^{-4}$.
On the other hand,
for the high X-ray flux case of {\it Chandra} (MJD 55456) observation,
we obtain 
\begin{equation}
  n_{\rm s,obs} 
  = 4.6 \times 10^7 e^{-E/E_{\rm br}} (r/6r_{\rm g})^{-2} 
    \mbox{ cm}^{-3}.
  \label{eq:n_gamma_obs_2}
\end{equation}
Therefore, we find $E_{\rm br}<140$~keV. 

The temperature of Comptonizing electrons is estimated to be
$T_{\rm e}= 2 E_{\rm br}$ for an optically thin plasma and
$T_{\rm e}= 3 E_{\rm br}$ for an optically thick one
\citep{petrucci10}.
Thus, we obtain the conservative upper bound, 
$T_{\rm e}<1.2$~MeV and $T_{\rm e}<420$~keV
for the low and high X-ray flux cases, respectively.
Considering that the Compton optical thickness is probably
thin at small $\dot{m}$, 
$T_{\rm e}<740$~keV and $T_{\rm e}<280$~keV
may be more appropriate as the upper bound.

Adopting an ADAF theory \citep{mahad97}, 
one finds $T_{\rm e} < 600$~keV at $\dot{m} > 10^{-4}$.
Therefore, the density of the MeV photon emitted by ADAF
during the jet phase (not during the VHE flare phase)
appear to be consistent with the typical MeV photon density
deduced from the X-ray observations during the low X-ray flux phase.
However, if we adopt the high X-ray flux phase,
the required electron temperature, $T_{\rm e}<280$~keV,
appears to be too low.
In addition, the e-folding energy
is empirically deduced to be above 150~keV 
(i.e., $T_{\rm e}>300$~keV is obtained)
commonly in low accretion rate systems 
(\citealp{santo12} for galactic black holes;
 \citealp{nowak11} for Cyg~X-1;
 \citealp{maliz14} for Seyfert 1s).
Since $T_{\rm e}$, and hence $E_{\rm br}$ increases 
with decreasing $\dot{m}$,
the MeV photon density may increase with decreasing $\dot{m}$ 
when $\dot{m} \ll 10^{-4}$,
although the normalization of the photon number density
decreases in X-rays (e.g., in $2-10$~keV).
On these grounds, 
our ansatz of dropping the accretion rate by half 
in order to create a BH gap, will be invalidated, 
if X-ray observations indicate a MeV photon density that exceeds 
equation~(\ref{eq:n_gamma}) for IC~310.

What is more, we used $r=6r_{\rm g}$ to evaluate $n_{\rm s}$
and $n_{\rm s,obs}$ in this subsection, 
because the lower-limit radius of the ADAF emission is set to be
$r_{\rm min}=6r_{\rm g}$ in the self-similar analytical solution
we are adopting 
(\citealp{mahad97}; see also e.g., \citealp{narayan95b,lasota96}
 for an accordant choice of $r_{\rm min}$).
Note that the radius $r=6r_{\rm g}$ gives a conservative estimate
(i.e., in this case, higher value) of $n_{\rm s,obs}(r)$ 
than $r=10r_{\rm g}$ does
(see the third paragraph of \S~\ref{sec:Lf}).
It is, however, possible to compute the RIAF (including ADAF)
down to the horizon, 
if we solve the hydrodynamical equations 
\citep{manmoto00,li08}
or the magnetohydrodynamical equations
\citep{punsly09,mckinney12}
in the Kerr spacetime.
For instance, $n_{\rm s,obs}$ may increase inwards with $\propto r^{-2}$
also in $r<6r_{\rm g}$; 
in this (simplified) case, we would underestimate $n_{\rm s,obs}$ 
at the gap center ($r=2r_{\rm g}$) by nine times.
An underestimate of $n_{\rm s,obs}$, in turn, 
incurs an optimistic evaluation of the
charge-starvation condition, $n_{\rm s,obs} < n_{\rm GJ}$,
although the $r$ dependence of $B$ (and hence that of $n_{\rm GJ}$) 
is non-trivial.
If an observation shows that the MeV photon density 
gives $n_{\rm s,obs} > n_{\rm GJ}$,
gaps cannot be formed around the BH during the period.
To constrain the pair-creating photon density more conclusively,
simultaneous observations in TeV and MeV energies are crucial.

\subsection{Null surface versus stagnation surface}
\label{sec:position}
As noted in \S~\ref{sec:gap_position},
the gap center shift outwards
if electrons are injected across the inner boundary.
In particular, figure~\ref{fig:rho_5} shows that
the gap center will shift to $r > 6 r_{\rm g}$ 
if the injected current attains $\approx \Omega B/(2\pi)$.
Since a low plasma density is expected around the stagnation surface, 
it is reasonable to consider a gap formation around it 
under the existence of such a substantial current injection
across the inner boundary. 

From the analogy with the pulsar outer gap model \citep{hiro01b},
we can expect that the BH gap electrodynamics is little affected
by the change of the gap position.
This is because the gap solution is essentially 
described by the factor $w/l_{\rm GJ}$ (e.g., eq.~[\ref{eq:Lgap}]),
where $l_{\rm GJ}$ denotes the length scale in which $\rho_{\rm GJ}$
changes substantially.
If the gap is located around the null surface,
we can put $l_{\rm GJ} \approx r_{\rm H}$;
this dimension analysis would give the similar results 
as we have derived solving the spatial dependence of $\rho_{\rm GJ}$
explicitly.
On the contrary, 
if the gap is located around the stagnation surface
at radius $r$, we can put $l_{\rm GJ} \approx r$.
Thus, if the gap extends enough and $w \sim r$ holds,
the potential drop approaches the EMF exerted across the horizon
and $L_{\rm gap}$ approaches $L_{\rm sd}$,
again in the same manner as pulsars.
Thus, the results will not change dramatically
if the gap center shifts from the null surface
to another place, such as the stagnation surface.

In \S~\ref{sec:gap_position}, we point out the two representative
positions for a gap to be formed: 
the null surface and the stagnation surface.
It is, however, worth noting that more than one gap will not be formed
along the same magnetic field lines.
This is because the pairs cascaded outside one gap will migrate
into the other gap to screen the $E_\parallel$ there when they polarize.
That is, an injection of pairs quenches the gap when one of the charges
(e.g., electrons) return, whereas  
an injection of a completely-charge-separated flow, 
which consists of only one sign of charges (e.g., positrons), 
merely shifts the gap position.
In the present solutions,
the gap around the null surface is {\it not}
quenching another gap that would arise at another place.
The Poisson equation (\ref{eq:pois}) merely does not allow
a gap to be formed around other places than the null surface,
if there is no current injection 
across either the inner or the outer boundaries.
In subsequent papers, we will examine a gap formation at another 
positions (e.g., around the separation surface),
considering a current injection cross the inner boundary.

\subsection{Comparison with other BH gap models}
\label{sec:comparison}
In the present paper, we applied our model to IC~310,
while LR11 and BT15 to M87 and SgrA*.
Apart from the applied objects, 
there are important electrodynamical differences 
among the three works.
We discuss this point in this section.

\subsubsection{Comparison with LR11}
\label{sec:comparison_LR}
Let us compare the present work with LR11.
The main difference appears in two points:
The dominant emission process and the gap closure condition.

First, we consider the dominant emission process.
To demonstrate the formation of a power-law-like spectrum in VHE,
LR11 assumed that the IC process takes place in the Thomson regime
and evaluated $\gamma_{\rm IC}$ by
\begin{equation}
  \gamma_{\rm Thomson}= \sqrt{\frac{e E_\parallel}
                                 {\sigma_{\rm T} u_{\rm s}}
                           }, 
  \label{eq:Thomson}
\end{equation}
where $\sigma_{\rm T}$ denotes the Thomson cross section
and $u_{\rm s}$ the soft photon energy density.
We should notice here that only the soft photons having energies below
$E_{\rm s}= m_{\rm e}c^2/\gamma$
contribute in the Thomson regime,
unless collisions take place with tiny angles.
To compute $u_{\rm s}$, 
we have to integrate the ADAF photon density up to the energy $E_{\rm s}$.
Therefore, if $E_{\rm s}$ appears much less than the ADAF peak energy,
we obtain $u_{\rm s} \ll u_{\rm submm}$,
where $u_{\rm submm}$ denotes the ADAF photon density in
submillimeter wavelengths and essentially represent the soft photon
density integrated over all the photon energies.

In the present case, $\gamma > 3 \times 10^9$ 
(solid curve in the middle panel of figure~\ref{fig:results}) 
means that only the ADAF photons with energies 
$E_{\rm s} < 2 \times 10^{-4}$~eV
contribute in Thomson regime.
Since the ADAF spectrum peaks between $10^{-3}$~eV and $10^{-2}$~eV,
we thus obtain $u_{\rm s} \ll u_{\rm submm}$.
The thin dotted curve in the middle panel of figure~\ref{fig:results}
is computed with this reduced $u_{\rm s}$. 
On the other hand, 
LR used $u_{\rm submm}$ to evaluate equation~(\ref{eq:Thomson}),
underestimating $\gamma_{\rm IC}$ such that 
$\gamma_{\rm IC} < \gamma_{\rm curv}$, or equivalently,
the the IC process dominates the curvature process.

In short,
the leptonic Lorentz factors are turned out to be regulated
via the curvature process in the present paper, 
while LR11 considered that they are regulated via the IC process.
To compute the IC-limited Lorentz factors,
which is essential to examine the closure condition,
we employ the Klein-Nishina cross section 
taking account of the ADAF spectrum, 
while LR11 employed the Thomson limit
using a single parameter $u_{\rm submm}$.

Second, we compare the gap closure condition.
LR11 considered a necessary condition for a BH gap to be sustained
and imposed that the cascaded pair density should exceed
the GJ value outside the gap.
They proposed the closure condition,
${\cal M}_{\rm LR} > n_{\rm pair} / (\vert \rho_{\rm GJ} \vert / e)$,
where ${\cal M}_{\rm LR}$ denotes the multiplicity of cascaded pairs 
from a sing primary lepton (their eq.~[25]).
However, as demonstrated in the bottom panel of figure~\ref{fig:results},
the cascaded pair density always exceeds the GJ value
(horizontal dotted line).
Thus, our solutions automatically satisfy the necessary condition
proposed by LR11.

Third, we discuss minor differences.
For one thing, we consider the gap arising around the
null surface, while LR11 around the stagnation surface.
Nevertheless, it will not incur a qualitative difference
as demonstrated in the pulsar outer gap model \citep{hiro01b}.
What is more,
our results show that the $\gamma$-ray luminosity is
proportional to $(w/r_{\rm g})^4$ while LR11 $(w/r_{\rm g})^2$;
the additional $(w/r_{\rm g})^2$ dependence comes from the fact that
the electric potential drop in the gap is proportional to $w$
and that the maximum current is also roughly proportional to $w$
(eq.~[\ref{eq:Jmax}]).
For pulsar out-gap models, the gap is geometrically thin
in the {\it transverse} direction of the magnetic field; 
thus, the potential drop is proportional to $(w_\perp/r_{\rm g})^2$,
where $w_\perp$ refers to the trans-magnetic-field thickness
of the gap.
As a result, the gap luminosity is proportional to 
$(w_\perp/r_{\rm g})^3$ \citep{hiro08},
because the total current flowing in the gap is proportional to
the gap cross section, which is proportional to $w_\perp$.
However, a BH gap is thin in the {\it longitudinal} direction
of the magnetic field;
thus, the potential drop is proportional to $(w/r_{\rm g})^3$,
which leads to the $(w/r_{\rm g})^4$ dependence of the gap luminosity
(eq.~[\ref{eq:Lgap}]).
One final point is that
we explicitly computed the photon energy dependence of 
the pair creation rate, ICS emissivity, and curvature emissivity,
while LR11 adopted a monochromatic approximation for the photon
specific intensity.

\subsubsection{Comparison with BT15}
\label{sec:comparison_BT}
Let us next compare the electrodynamics of this paper
with that of BT15.
In the latter paper, they neglected the $\rho_{\rm GJ}$ term
in the Poisson equation, their equation~(100).
We interpret that they assumed a very large created charge density
in the gap, that is, $\vert \rho \vert \gg \vert \rho_{\rm GJ} \vert$.
The reasons are as follows:
If the created charge density were sub-GJ,
$\vert E_\parallel \vert$ would become less than the vacuum case,
leading to 
$\vert E_\parallel \vert < E_\perp (w/l_{\rm GJ})^2$,
where $E_\perp = \Omega_{\rm F} \varpi B / c$
(i.e., their eq.~[1]),
$\varpi$ denotes the distance from the rotation axis,
$w$ corresponds to $\Delta$ in their notation, and
$l_{\rm GJ}$ the distance in which the GJ charge density changes
substantially.
In the case of BT15, they assume
that a gap is located at the stagnation surface, 
namely between $r \approx 5 r_{\rm g}$ and $10 r_{\rm g}$;
thus, $l_{\rm GJ} > 5 r_{\rm g}$,
which becomes $\approx 5 \times 10^{15}$~cm for M87.
It follows that their solution of $\Delta \approx 4.3 \times 10^{11}$~cm
(their eq.~[57]) gives 
$E_\parallel \approx \times 10^{-8} E_\perp$,
about 100 million times smaller than their $E=E_\perp$ 
(their eq.~[1]),
as long as the created charge is sub-GJ
as sketched in figure~\ref{fig:rho_5}.
To overcome difficulty, one must resort on a super-GJ
charge density created in the gap, $\rho \gg \rho_{\rm GJ}$.
For example, if the real charge density is $\sim 10^8$ times
greater than the GJ value,
it appears that the Poisson equation might give
$E_\parallel \sim E_\perp$, that is, their equation~(1),
although $\Delta/l_{\rm GJ} \sim 10^{-4}$.

\begin{figure}
 \epsscale{1.0}
 \plotone{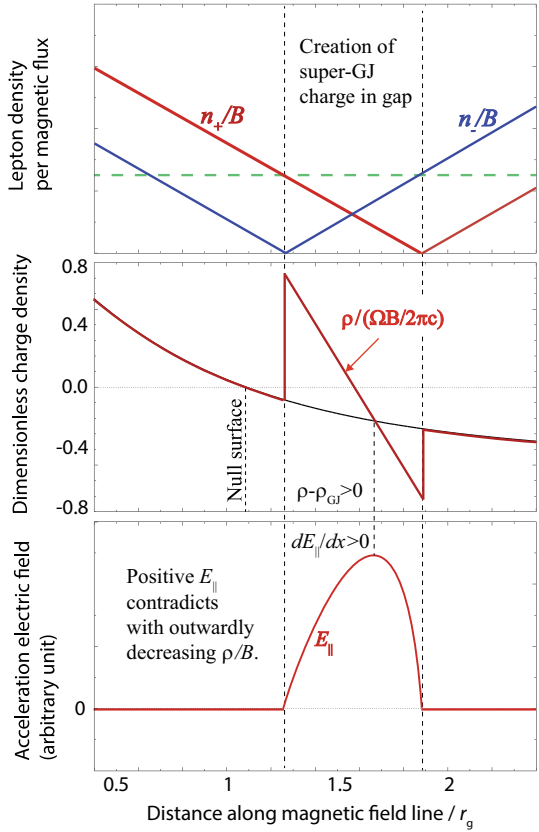}
\caption{
Similar figure as figure~\ref{fig:rho_2},
but the created current density is super-Goldreich-Julian.
The gap center is assumed to be remotely located
from the null surface.
\label{fig:rho_6}
}
\end{figure}

If the created charge density were super GJ, 
it would distribute as the red solid line 
in the middle panel of figure~\ref{fig:rho_6}.
However, in this case, $\rho-\rho_{\rm GJ}$ is positive
(or negative) in the inner (or outer) part of the gap,
leading to a positive $E_\parallel$,
which in tern contradicts with the outwardly decreasing $\rho$.
On the contrary, if $\rho$ increases outward, 
a negative (or positive) $\rho-\rho_{\rm GJ}$ in the inner (or outer)
part of the gap, would lead to a negative $E_\parallel$,
which contradicts with the outwardly increasing $\rho$.
Thus, we consider that a super-GJ charge creation within the gap
is implausible.
This is, indeed, the physical reason why all the pulsar gap models
treat sub-GJ charge creation within the gap.

On these grounds, we propose to replace their equation~(1) with 
$\vert E_\parallel \vert \approx E_\perp (w/l_{\rm GJ})^2$,
which approximates our equation~(\ref{eq:Epara}).
In addition, we propose to replace their equation~(17) with 
our condition ${\cal M}=1$ (eq.~[\ref{eq:MP}]), or with
${\cal M}_{\rm LR} > n_\pm / (\vert \rho_{\rm GJ} \vert / e)$
of LR11.

\subsubsection{Comparison with HO98}
\label{sec:comparison_HO}
In HO98, they solved a set of Poisson equation,
Boltzmann equations for electrons and positrons, and
the radiative transfer equation for a supermassive BH parameter.
They considered an application to quasars and adopted 
a large accretion rate,
$\dot{m} \sim 0.1$.
It was demonstrated that a BH gap supply sufficient pairs
so that a force-free magnetosphere may be sustained outside the gap,
and that the gap luminosity increases with decreasing accretion rate,
or the soft photon density.
In addition, they concluded that the BH gap luminosity is totally
undetectable, because the gap width is very thin,
$w<0.02 r_{\rm H}$ if $\dot{m} \sim 0.1$.
However, they did not notice that a BH-gap emission would be detectable
if accretion rate reduces sufficiently.

In the present paper, we therefore consider 
low luminosity active galactic nuclei and adopted 
$\dot{m} < 10^{-4}$.
As the first paper in this series,
we only analytically examined the gap closure condition,
which is proved to be valid both in BH magnetospheres (HO98)
and in pulsar magnetospheres \citep{hiro13}
through the comparison with the numerical solutions of the set of 
Maxwell-Boltzmann equations.

\subsection{Comparison with pulsar outer-gap model}
\label{sec:comparison_PSR}
Let us compare the present BH gap model
with the pulsar outer-magnetospheric particle accelerator model,
that is, the outer-gap (OG) model.

First, in a pulsar magnetosphere, the null surface appears because of the
convex geometry of a dipole magnetic field.
On the other hand, in a BH magnetosphere,
a null surface appears due to the frame-dragging effect 
around a rotating black hole.
Thus, for the same magnetic field polarity,
the sign of $E_\parallel$ is opposite.
In a BH magnetosphere, a pulsar-like null surface could also appear
far away from the horizon, in principle.
However, unless a strong ring current flows within a certain radius,
a dipolar field will not be formed in a BH magnetosphere.
Various numerical simulations
\citep{hirose04,mckinney06,mckinney07a,tchek10,punsly09,mckinney12}
seems to rule out the formation of such a dipolar-like field geometry,
except for the initial-condition-dependent closed poloidal field lines
in the equatorial torus imposed far away from the horizon.

Second, in the OG model, soft photon field is provided by
the cooling neutron star (NS) thermal emission and/or the heated 
polar-cap thermal emission.
The NS surface emissions peak in the X-ray energies.
Thus, the curvature 1-10~GeV photons efficiently materialize as pairs
within the gap, thereby contributing for the gap closure.
For very young pulsars like the Crab pulsar,
IC photons could also materialize.
However, the thermal IR photon field is much weaker 
than the thermal X-ray field;
thus, the IC photons do not contribute for gap closure in pulsars.
In the BH gap model,
it is the ADAF that provides the soft photon field.
The ADAF spectrum peaks in submillimeter wavelengths.
Thus, the 100~TeV IC photons efficiently materialize as pairs
to close the gap.
The 0.1-10~TeV curvature photons, on the other hand,
do not materialize efficiently,
because the IR photon number flux is 5-6 orders of magnitude weaker
than the submillimeter one.

Third, if a NS is isolated or 
is not accreting plasmas from the companion in a binary system,
the magnetosphere becomes highly vacuum. 
The pair density becomes comparable to the GJ value 
if the space-charge-limited flow is drawn from the NS surface,
or becomes much less than the GJ charge value if there is no such
a plasma injection.
Therefore, a gap is inevitably formed to replenish
the charge-starved magnetosphere with a large duty cycle,
which is essentially unity.
In a black hole magnetosphere, on the other hand,
ADAF can supply enough charges not only in the lower latitudes
but also in the polar funnel via photon-photon pair production.
In the case of IC~310, the BH gap is, therefore, expected to be
intermittent, because the BH gap can be turned on only when
the accretion rate becomes less than half of the time-averaged value.

Fourth, 
a NS magnetosphere is formed by the electric currents inside the NS,
whereas a BH magnetosphere is formed by those outside the BH.
Thus, the magnetic field energy dominates the plasma energy 
in a NS magnetosphere,
whereas it is at most comparable to the plasma's gravitational binding energy 
in a BH magnetosphere.
Thus, an accreting NS allows accretion along the 
higher-latitude magnetic field lines that would be open without accretion.
This enhanced plasma density quenches an OG, 
once plasma accretion takes place in a binary system.
On the contrary, an accreting BH allows accretion only in the
lower latitudes (i.e., near the equator with a certain vertical height),
prohibiting plasma penetration into the funnel region,
because the centrifugal-force barrier prevents plasma accretion 
towards the rotation axis.
Thus, even under plasma accretion, a BH gap can be switched on,
provided that the ADAF-supplied pair density (via photon-photon collisions)
is small compared to the GJ value.

Fifth and finally,
OG distribution is highly non-axisymmetric
for an oblique rotator (i.e., a pulsar).
This is because $\rho_{\rm GJ}$ and hence the null surface distributes
highly non-axisymmetrically,
reflecting the three-dimensional magnetic field geometry.
However, $\rho_{\rm GJ}$ and hence the gap distributes 
nearly spherically in a BH magnetosphere, 
irrespective of the magnetic field geometry
(or the magnetic inclination angle), 
as long as $\Omega_{\rm F}$ is nearly constant for all the field lines.
Therefore, the BH gap solution, which is governed
by the $\rho_{\rm GJ}$ distribution, little depends on the
magnetic field geometry (e.g., split-monopole or parabolic),
because $\rho_{\rm GJ}$ distribution is essentially determined
by the frame-dragging effect near the horizon,
instead of the functional form of $A_{\rm \varphi}(r,\theta)$,
particularly in the higher latitudes
(fig.~\ref{fig:side_null}).
 
In addition to the differences point out above,
there are, of course, similarities between the present BH gap model
and the OG model, 
because the basic equations are exactly the same.

The first and foremost,
the gap width, and hence the 
$\gamma$-ray efficiency,
which is defined by the ratio between the
$\gamma$-ray and spin-down luminosities, 
increases with decreasing soft photon flux.
This is because the pair-production mean-free path
increases with decreasing photon flux.
For instance, the OG $\gamma$-ray efficiency increases with
decreasing NS surface emission,
or equivalently with increasing NS age \citep{hiro13}.
In the same manner, the BH gap $\gamma$-ray efficiency
increases with decreasing accretion rate of ADAF.

Next, superposition of the curvature radiation
from different positions produces a nearly power-law
spectrum in a limited energy range.
In a pulsar magnetosphere, the caustic effect results a compilation of
emission from different altitudes
with varying $\rho_{\rm c}$ and $E_\parallel$.
However, in a BH magnetosphere, 
highly tangled, instantaneous magnetic field lines 
possibly result in a superposition of the curvature emission from
different places with varying $\rho_{\rm c}$ and $E_\parallel$.

\subsection{General relativistic effects on the accretion flow}
\label{sec:GR_acc}
In the present paper, we have employed the Newtonian, self-similar,
analytical solution of the ADAF \citep{mahad97}
to specify the background, 
soft photon field that illuminates the gap.
What is more, for simplicity,
we have assumed that the soft photon field is homogeneous
within $10r_{\rm g}$ and its flux reduces by $r^{-2}$ law
outside this radius,
and that the collision angles between these soft photons and
the hard $\gamma$-rays are isotropic.
However, in a realistic BH magnetosphere,
position-dependent specific intensity of the
soft photon field is necessary to further quantify the
photon-photon absorption as well as the IC scatterings.
Thus, the next step may be to incorporate 
the general-relativistic calculations of radiatively inefficient
accretion flows (RIAF). 
Since the photon distribution is found to be highly 
inhomogeneous and anisotropic around a rapidly rotating BH
\citep{li09}, 
we may be able to constrain the spin of the individual BHs,
if we combine the present method, which specifies the
emissivity distribution in the magnetosphere,
with a sophisticated, numerical RIAF theory.

\acknowledgments
One of the authors (K. H.) is indebted to Dr. D. Glawion 
for valuable discussion on the MAGIC data and to 
Dr. K. Mannheim for commenting on the draft. 
This work is supported by the Theoretical Institute for Advanced Research 
in Astrophysics (TIARA) operated under Academia Sinica, 
and the Formosa Program between National Science Council in Taiwan and 
Consejo Superior de Investigaciones Cientificas in Spain administered 
through grant number NSC100-2923-M-007-001-MY3.


\begin{thebibliography}{}



\bibitem[Aleksic et al.(2011)]{alek11} 
  Aleksi\`c, J. et al. 
  2010, \apj 723, L207

\bibitem[Aleksic et al.(2014a)]{alek14a} 
  Aleksi\`c, J. et al. 
  2014, \aap 563, A91

\bibitem[Aleksic et al.(2014b)]{alek14b} 
  Aleksi\`c, J. et al. 
  2014, Science 346, 1080

\bibitem[Bardeen (1970)]{bard70}
  Bardeed, J. M.
  1970, \nat 226, 64

\bibitem[Barkov et al.(2010)]{barkov10}
  Barkov, M. V., Aharonian, F. A., Bosch-Ramon, V. 
  2010, \apj 724, 1517 

\bibitem[Barkov et al.(2012)]{barkov12}
  Barkov, M. V., Bosch-Ramon, V. Aharonian, F. A., 
  2010, \apj 755, 170

\bibitem[Bednarek \& Protheroe (1997)]{bednarek97}
  Bednarek, W., Protheroe, R. J. 
  1997, \mnras 287,L9 

\bibitem[Beskin et al. (1992)]{bes92}
  Beskin, V.~S., Istomin, Ya.~N., \& Par'ev, V.~I. 
  1992, Sov. Astron. 36(6), 642

\bibitem[Beskin \& Zheltoukhov(2013)]{bes13}
  Beskin, V.~S., Zheltoukhov, A. A.
  2013, Astron. Lett., 39, 215

\bibitem[Blandford \& Znajek (1977)]{bla76}
  Blandford, R.~D., \& Znajek, R.~L. 
  1976, \mnras, 179, 433

\bibitem[Boyer \& Lindquist (1967)]{boyer67}
  Boyer, R. H. \& Lindquist, R. W.
  1967 {\it J. Math. Phys.} 265, 281

\bibitem[Broderick \& Tchekhovskoy (2011)]{brod15}
  Broderick, A.~E., Tchekhovskoy A.
  2015, ApJ 809, 97

\bibitem[Camenzind(1986a)]{came86a} 
  Camenzind, M. A.
  1986a, \aap, 156, 137

\bibitem[Camenzind(1986b)]{came86b} 
  Camenzind, M. A.
  1986b, \aap, 162, 32

\bibitem[Cheng et al.(1986a)]{cheng86a} 
  Cheng, K. S., Ho, C. \&  Ruderman, M.
  1986, \apj, 300, 500


\bibitem[Cheng et al.(2000)]{cheng00} 
  Cheng, K. S., Ruderman, M. \& Zhang, L.
  2000, \apj, 537, 964

\bibitem[Del Santo(2012)]{santo12} 
  Del Santo, M. 
  2012, http://arxiv.org/abs/1209.2880

\bibitem[Eisenacher et al.(2013)]{eisen13} 
  Eisenacher, D. et al., 
  2013, Proc. to the 33rd ICRC, Id. 0336, Rio de Janerio, Brazil

\bibitem[Schulz, Kadler \& Ros et al.(2015)]{schulz15} 
  Schulz, R., Kadler, M., Ros, E., Eisenacher Glawion D., 
  Bach, U., Els\"{a}sser, D., Grossberger, C.,
  Kreykenbohm, I., Mannheim, K., M\"{u}ller, C.,
  Tr\"{u}stedt, J., Wilms, J.
  arXiv:1502.03559

\bibitem[Giannios et al. (2010)]{giannios10}
  Giannios, D., Uzdensky, D. A., Begelman, M. C. 
  2010, \mnras 402, 1649 

\bibitem[Ghisellini et al. (2014)]{ghise14}
  Ghisellini, G. et al.
  2014, \nat 515, 376

\bibitem[Goldreich \& Julian (1969)]{GJ69}
  Goldreich, P., Julian, W. H.,
  1969, \apj 157, 869

\bibitem[Hirose et al. (2004)]{hirose04}
  Hirose, S. Krolik, J., de Villiers, J.-P., Hawley, J. F.
  0224, \apj 606, 1083

\bibitem[Hirotani et al. (1992)]{hiro92}
  Hirotani, K. Takahashi, M., Nitta, S., Tomimatsu, A.
  1992, \apj 386, 455

\bibitem[Hirotani et al. (1993)]{hiro93}
  Hirotani, K., Tomimatsu, A., Takahashi, M.
  1993, \pasj 45, 431

\bibitem[Hirotani \& Okamoto (1998)]{hiro98}
  Hirotani, K. \& Okamoto, I.
  1998, \apj, 497, 563


\bibitem[Hirotani \& Shibata (1999b)]{hiro99}
  Hirotani, K., \& Shibata, S. 
  1999, \mnras 308, 67

\bibitem[Hirotani \& Shibata (2001)]{hiro01a}
  Hirotani, K., \& Shibata, S. 
  2001, \apj 558, 216

\bibitem[Hirotani \& Shibata (2001)]{hiro01b}
  Hirotani, K., \& Shibata, S. 
  2001, \mnras 325, 1228

\bibitem[Hirotani et al.(2003)]{hiro03}
    Hirotani, K., Harding, A. K., \& Shibata, S.,
    2003, \apj 591, 334

\bibitem[Hirotani(2008)]{hiro08}
    Hirotani, K.
    2008, \apj 688, L25

\bibitem[Hirotani(2006)]{hiro06} 
  Hirotani, K.
  2006b, Mod. Phys. Lett. A (Brief Review) 21, 1319 

\bibitem[Hirotani (2011)]{hiro11}
  Hirotani, K. 
  2011, \apj 733, L49

\bibitem[Hirotani (2013)]{hiro13}
  Hirotani, K.
  2013, \apj 766, 98

\bibitem[Hirotani(2015)]{hiro15} 
  Hirotani, K.
  2015, \apj 798, L40

\bibitem[Ichimaru (1977)]{ichimaru77}
  Ichimaru, S.
  1979, \apj 214, 840

\bibitem[Jauch \& Rohrlich (1955)]{jauch55}
  Jauch, J. M., \& Rohrlich, F.
  1955, {\it Theory of photons and electrons}
  (Addison-Wesley Publishing Inc., Massachusetts)

\bibitem[Kerr (1963)]{kerr63}
  Kerr, R. P.
  \prl 11, 237

\bibitem[Koide et al. (2002)]{koide02}
  Koide, S. et al.
  1976, \mnras, 179, 433

\bibitem[Krolik et al. (2005)]{krolik05}
  Krolik, J. H., Hawley, J. F., Hirose, S.
  2005, \apj, 622, 1008

\bibitem[Lasota et al. (1996)]{lasota96}
  Lasota, J. P., Abramowicz, M. A., Chen, X., Krolik, J., 
  Narayan, R., \& Yi, I.
  1996, \apj 462, 192

\bibitem[Levinson \& Rieger (2011)]{levi11}
  Levinson, A., Rieger, F.
  2011, \apj 730, 123

\bibitem[Li et al. (2008)]{li08}
  Li, Y.-R., Yuan, Y.-F., Wang, J.-M., Wang, J.-C., Zhang, S.
  2008, \apj 699, 513

\bibitem[Mahadevan (1997)]{mahad97}
  Mahadevan, R.
  1997, \apj 477, 585

\bibitem[Malizia et al.(2014)]{maliz14}
  Malizia, A., Molina, M., Bassani, L., Stephen, J. B., 
  Bazzano, A., Ubertini, P., Bird, A. J.
  2014, \apj 782, L25

\bibitem[Manmoto et al.(1997)]{manmoto97}
  Manmoto, T., Mineshige, S., Kusunose, M.
  1997, \apj 489, 791

\bibitem[Manmoto (2000)]{manmoto00}
  Manmoto, T.
  2000, \apj 534, 734

\bibitem[Merloni, Heinz \& di Matteo (2003)]{merloni03}
  Merloni, A., Heinz, S., di Matteo, T. 
  2003, \mnras 345, 1057

\bibitem[McElroy (1995)]{macElroy95}
  McElroy, D.~B., 
  1995, \apjs 100, 105

\bibitem[McKinney et al.(2006)]{mckinney06}
  McKinney, J.~C.
  2006, \mnras 368, 1561

\bibitem[McKinney et al.(2007a)]{mckinney07a}
  McKinney, J.~C., Narayan, R.
  2007a, \mnras 375, 513

\bibitem[McKinney et al.(2007b)]{mckinney07b}
  McKinney, J.~C., Narayan, R.
  2007b, \mnras 375, 531

\bibitem[McKinney et al.(2012)]{mckinney12}
  McKinney, J.~C., Tchekhovskoy, A., Blandford, R.~D.
  2012, \mnras 423, 3083

\bibitem[Mestel (1971)]{mestel71}
  Mestel L. 
  1971, Nature 233, 149

\bibitem[Michel(1973)]{michel73}
  Michel F. C. 
  1973, \apj 180, L133

\bibitem[Nakamura et al.(1997)]{nakamura97}
  Nakamura, K. E., Matsumoto, R. Kusunose, M., Kato, S.
  1997, \pasj 49, 503

\bibitem[Narayan \& Yi (1994)]{narayan94} 
  Narayan, R., Yi, I.
  1994, \apj 428, L13

\bibitem[Narayan \& Yi (1995)]{narayan95} 
  Narayan, R., Yi, I.
  1995, \apj 452, 710

\bibitem[Narayan et al.(1995)]{narayan95b} 
  Narayan, R., Yi, I., Mahadevan, R.
  1995, \nat 374, 623

\bibitem[Narayan et al.(1997)]{narayan97} 
  Narayan, R., Kato, S., Honma, F.
  1997, \apj 476, 49

\bibitem[Neronov et al.(2010)]{nero10} 
  Neronov, A., Semikov, D., Vovk, I. 
  2010, \aap 519, L6

\bibitem[Nowak et al.(2011)]{nowak11} 
  Nowak, M. A., Hanke, M., Trowbridge, S. N., Markoff, S. B.,
  Wilms, J., Pottschmidt, K., Coppi, P., Maitra, D., 
  Davis, J. E.; Tramper, F.
  2011, \apj 728, 13

\bibitem[Petrucci et al.(2010)]{petrucci10} 
 Petrucci, P. O., Haardt, F. Maraschi, L. et al.
 2010, \apj 721, 1340

\bibitem[Phinney(1983)]{phinney83} 
 Phinney, S. 1983, unpublished PdH thesis, Cambridge University

\bibitem[Punsly et al.(2009)]{punsly09} 
 Punsly, B., Igumenshchev, I. V., Hirose, S.
 2009, \apj 704, 1065

\bibitem[Punsly (2011)]{punsly11} 
 Punsly, B.
 2011, \mnras 418, L138



\bibitem[Rhee, Burns \& Kowalsky(1994)]{rhee94} 
 Rhee, G., Burns, J. O., Kowalski, M. P.
 1994, \aj 108, 1137

\bibitem[Romani (1996)]{romani96}
  Romani, R. 
  1996, \apj 470, 469

\bibitem[Li et al.(2009)]{li09}
  Li, R.-Y., Yuan, Y.-F., Wang, J.-M., Wang, J.-C., Zhang, S.
  2009, \apj 513, 524

\bibitem[Rybicki \& Lightman (1979)]{rybicki79} 
  Rybicki, G. B. \& Lightman, A. P.
  1979, Radiative Processes in Astrophysics
 (New York: John Wiley \& Suns)

\bibitem[Sato et al.(2005)]{sato05} 
  Sato, K. et al., 
  2005, \pasj 57, 743

\bibitem[Schwarz et al.(1992)]{schwarz92} 
  Schwarz, R. A. et al. 
  1992, \aap 256, L11

\bibitem[Sijbring \& de~Bruyn (1998)]{sijb98}
  Sijbring, D., \& de~Bruyn, A.~G.
  1998, \aap 331, 901

\bibitem[Sikora \& Begelman(2013)]{sikora13}
  Sikora, M., Begelman, M.~C.
  2013, \apj 764, L24 

\bibitem[Simien \& Prugniel (2002)]{simien02}
  Simien, F., Prugniel, P.
  2002, \aap 384, 371

\bibitem[Spitkovski(2006)]{spit06} %
  Spitkovsky, A.
  2006, \apj 648, L51

\bibitem[Takahasi et al.(1990)]{taka90}
  Takahashi, M., Nitta, S., Tatematsu, Y., Tomimatsu, A.
  1990, \apj 363, 206

\bibitem[Tchekhovskoy et al.(2010)]{tchek10}
  Tchekhovskoy, A., Narayan, R., McKinney, J.~C.
  2010, \apj 711, 50

\bibitem[Thorne (1974)]{thorne74}
  Thorne, K. S.
  1974, \apj 191, 507


%














\end{thebibliography}
\end{document}